\documentclass[useAMS,usenatbib,referee]{mn2e}

\usepackage{amsmath, amssymb}
\usepackage[dvips]{graphicx}
\usepackage{epsfig}

\newcommand{\grad}{\mathbf{\nabla}}
\newcommand{\rot}{\mathbf{\nabla} \times}
\newcommand{\divg}{\mathbf{\nabla}\cdot}

\newcommand{\rlight}{r_{\rm L}}
\newcommand{\Rs}{R_{\rm s}}
\newcommand{\Lsd}{L_{\rm sd}}

\newcommand{\er}{\mathbf{e}_{\rm r}}
\newcommand{\etheta}{\mathbf{e}_\vartheta}
\newcommand{\ephi}{\mathbf{e}_\varphi}

\newcommand{\mnras}{MNRAS}
\newcommand{\apss}{Ap\&SS}

\newcommand{\apj}{ApJ}

\newcommand{\prd}{Physical Review D}

\title[GREM fields around a slowly rotating neutron star]{General-relativistic electromagnetic fields around a slowly rotating neutron star: stationary vacuum solutions} 
\author[J. P\'etri]{J.  P\'etri$^{1}$
\thanks{E-mail: jerome.petri@astro.unistra.fr} \\
  $^{1}$Observatoire Astronomique de Strasbourg, Universit\'e de Strasbourg, CNRS, UMR 7550, 11 rue de l'Universit\'e, 67000 Strasbourg, France.}

\begin{document}

\date{Accepted . Received ; in original form }

\pagerange{\pageref{firstpage}--\pageref{lastpage}} 
\pubyear{2013}

\maketitle

\label{firstpage}

\begin{abstract}
  Pulsars are thought to be highly magnetized rotating neutron stars accelerating charged particles along magnetic field lines in their magnetosphere and visible as pulsed emission from the radio wavelength up to high energy X-rays and gamma-rays. Being highly compact objects with compactness close to $\Xi = R_s/R\approx0.5$, where $\Rs=2\,G\,M/c^2$ is the Schwarzschild radius and $\{M,R\}$ the mass and radius of the neutron star, general-relativistic effects become important close to their surface. This is especially true for the polar caps where radio emission is supposed to emanate from, leading to well defined signatures such as linear and circular polarization. In this paper, we derive a general formalism to extend to general relativity the Deutsch field solution valid in vacuum space. Thanks to a vector spherical harmonic expansion of the electromagnetic field, we are able to express the solution to any order in the spin parameter~$\Omega$ of the compact object. We hope this analysis to serve as a benchmark to test numerical codes used to compute black hole and neutron star magnetospheres.
\end{abstract}


\section{INTRODUCTION}

In our current understanding of pulsar magnetospheres and radiation mechanisms, strongly magnetized rotating neutron stars play a central role. The underlying plasma processes like particle acceleration, pair creation and pulsed emission profiles throughout the whole electromagnetic spectrum strongly depend on the peculiar magnetic field geometry and strength adopted or extracted from numerical simulations of the magnetosphere. For instance radio emission is believed to emanate from the polar caps, therefore in regions of strong gravity where curvature and frame-dragging effects are considerable due to the high compacity of neutron stars $\Xi = \Rs/R \approx0.5$ for typical models with its mass~$M$, its radius~$R$ and the Schwarzschild radius given by $\Rs=2\,G\,M/c^2$, $G$ being the gravitational constant and $c$ the speed of light. Detailed quantitative analysis of radio pulse polarization and pair cascade dynamics could greatly benefit from a better quantitative description of the electromagnetic field around the polar caps. Although there exists an extensive literature about flat space-time electrodynamics, only little work has been done to include general-relativistic effects.

The first general solution for an oblique rotator in flat vacuum space-time was found by~\cite{1955AnAp...18....1D} with closed analytical formulas. This solution is often quoted to explain the magnetic dipole radiation losses. To be truly exact, we emphasize that the Poynting flux~$\Lsd$ derived from his solution does not strictly coincide with the point dipole losses~$L_{\rm dipole}$ but depends on the ratio~$R/\rlight$, where $\rlight=c/\Omega$ is the light cylinder radius and $\Omega$ the rotation rate of the neutron star. It is only equal to the textbook equation for dipole losses in the limit of vanishing radius $\lim\limits_{R\to0} \Lsd = L_{\rm dipole}$. The distinction is meaningful at least for checking results emanating from numerical computations. Indeed, because of limited computer resources, we are often forced to take ratios~$R/\rlight\lesssim1$ not completely negligible compared to unity. Therefore the computed spin-down luminosity can significantly deviate from the point dipole losses. Moreover, \cite{1974AnPhy..87..244C} showed in the case of an aligned rotator that the electric field induced by frame-dragging effects could be as high as the one induced by the stellar rotation itself. These results were extended to an oblique rotator a few years later by~\cite{1980Ap&SS..70..295C} thanks to a formalism developed earlier by~\cite{1974PhLA...47..261C, 1974PhRvD..10.1070C, 1975PhLA...54....5C}. It is therefore crucial to treat Maxwell equations in the general-relativistic framework in order to analyse quantitatively acceleration and radiation in the vicinity of the neutron star. This led~\cite{1976GReGr...7..459P} to seek for an approximate solution of Maxwell equations in a curved space-time either described by the Schwarzschild metric or by the Kerr metric, using a linearised approach employing the Newman-Penrose formalism. He computed the structure of the electromagnetic waves propagating in vacuum and launched by a rotating dipole. He also gave an expression for the Poynting flux~$\dot{E}$ depending on the ratio~$R/\rlight$. The exact analytical solution for the static magnetic dipole in Schwarzschild space-time was given by \cite{1964ZETF...47..1030G, 1974PhRvD..10.3166P} and extended to multipoles by \cite{1970Ap&SS...9..146A}. 

\cite{1992MNRAS.255...61M} also studied the influence of space-time curvature and frame dragging effects on the electric field around the polar caps of a pulsar and confirmed the earlier claims of an increase in its strength. \cite{1995ApJ...449..224S} computed the electric field for an aligned rotator in vacuum in the Schwarzschild metric. The aligned rotator has also been investigated by \cite{2000PThPh.104.1117K} with special emphasize to particle acceleration in vacuum. \cite{1997ApJ...485..735M} and \cite{2003ApJ...584..427S} took a similar approach to study the acceleration of particles around polar caps. \cite{2001MNRAS.322..723R, 2002MNRAS.331..376Z, 2004MNRAS.352.1161R} computed the electromagnetic field in the exterior of a slowly rotating neutron star in the slow rotation metric as well as inside the star and investigated the impact of oscillations. They gave approximate analytical expressions for the external electromagnetic field close to the neutron star. \cite{2004MNRAS.348.1388K} extended the previous work by solving numerically the equations for the oblique rotator in vacuum in general relativity. They retrieve \cite{2001MNRAS.322..723R} results close to the surface and the Deutsch solution for distances larger than the light cylinder~$r\gg\rlight$.

It is the purpose of this paper to elucidate quantitatively and accurately some aspects of general-relativistic effects on the electrodynamics close to the neutron star. Our goal is to derive a general formalism to compute the solution of Maxwell equations in curved space-time for any multipole component of the magnetic field. Consequently, we use a 3+1 formalism of electrodynamics in curved space-time as presented in~\S\ref{sec:Modele}. Next we show how to solve for the electromagnetic field for an aligned rotator in~\S\ref{sec:Aligne}. This method is easily extended to a perpendicular rotator as explained in~\S\ref{sec:Orthogonal}. Because Maxwell equations in vacuum are linear, the most general solution for an oblique rotator will be a linear superposition of the weighted aligned and perpendicular rotator. Conclusions and future possible work are drawn in~\S\ref{sec:Conclusion}.

\section{The 3+1 formalism}
\label{sec:Modele}

The covariant form to describe the gravitational and electromagnetic field in general relativity is the natural way to write them down in a frame independent way. Nevertheless, it is more intuitive to split space-time into an absolute space and a universal time, similar to our all day three dimensional space, rather than to use the full four dimensional formalism. Another important advantage of a 3+1 split is a straightforward transcription of flat space techniques for scalar, vector and tensor fields to curved spaces. We start with a description of the special foliation used for the metric. Next we derive Maxwell equations in this foliation and conclude on some words about force-free electrodynamics which will be treated in another work but for completeness we give the useful expressions already in this paper.

\subsection{The split of the space-time metric}

We therefore split the four dimensional space-time into a 3+1~foliation such that the metric~$g_{ik}$ can be expressed as
\begin{equation}
  \label{eq:metrique}
  ds^2 = g_{ik} \, dx^i \, dx^k = \alpha^2 \, c^2 \, dt^2 - \gamma_{ab} \, ( dx^a + \beta^a \, c\,dt ) \, (dx^b + \beta^b \, c\,dt )
\end{equation}
where $x^i = (c\,t,x^a)$, $t$ is the time coordinate or universal time and $x^a$ some associated space coordinates. We use the Landau-Lifschitz convention for the metric signature given by $(+,-,-,-)$ \citep{LandauLifchitzTome2}. $\alpha$ is the lapse function, $\beta^a$ the shift vector and $\gamma_{ab}$ the spatial metric of absolute space.  By convention, latin letters from $a$ to $h$ are used for the components of vectors in absolute space (in the range~$\{1,2,3\}$) whereas latin letters starting from $i$ are used for four dimensional vectors and tensors (in the range~$\{0,1,2,3\}$). Our derivation of the 3+1 equations follow the method outlined by \cite{2011MNRAS.418L..94K}. A fiducial observer (FIDO) is defined by its 4-velocity~$n^i$ such that
\begin{subequations}
 \begin{align}
  n^i & = \frac{dx^i}{d\tau} = \frac{c}{\alpha} \, ( 1, - \mathbf \beta) \\
  n_i & = (\alpha \, c, \mathbf 0) 
 \end{align}
\end{subequations}
This vector is orthogonal to the hyper-surface of constant time coordinate~$\varSigma_t$. Its proper time~$\tau$ is measured according to
\begin{equation}
  d\tau = \alpha\,dt
\end{equation}
The relation between the determinants of the space-time metric~$g$ and the pure spatial metric~$\gamma$ is given by
\begin{equation}
  \sqrt{-g} = \alpha \, \sqrt{\gamma}
\end{equation}
For a slowly rotating neutron star, the lapse function is
\begin{equation}
  \label{eq:Lapse}
  \alpha = \sqrt{ 1 - \frac{R_s}{r} }
\end{equation}
and the shift vector
\begin{subequations}
 \begin{align}
  \label{eq:Shift}
  c \, \mathbf \beta = & - \omega \, r \, \sin\vartheta \, \ephi \\
  \omega = & \frac{R_s\,a\,c}{r^3}
 \end{align}
\end{subequations}
We use spherical coordinates~$(r,\vartheta,\varphi)$ and an orthonormal spatial basis~$(\er,\etheta,\ephi)$. The spin~$a$ is related to the angular momentum~$J$ by $J=M\,a\,c$. It follows that $a$ has units of a length and should satisfy $a \leq R_s/2$. Introducing the moment of inertia~$I$, we also have $J=I\,\Omega$. For the remainder of the paper, it is also convenient to introduce the relative rotation of the neutron star according to
\begin{equation}
 \tilde{\omega} = \Omega - \omega
\end{equation}
In the special case of a homogeneous and uniform neutron star interior with spherical symmetry, the moment of inertia is
\begin{equation}
  \label{eq:Inertie}
  I = \frac{2}{5} \, M \, R^2
\end{equation}
Thus the spin parameter can be expressed as
\begin{equation}
  \label{eq:spin}
  \frac{a}{R_s} = \frac{2}{5} \, \frac{R}{R_s} \, \frac{R}{\rlight}
\end{equation}
We adopt this simplification for the neutron star interior in order to compute the spin parameter~$a$.

\subsection{Maxwell equations}

Let $F^{ik}$ and ${^*F}^{ik}$ be the electromagnetic tensor and its dual respectively, see appendix~\ref{app:metric}. It is useful to introduce the following spatial vectors $(\mathbf B, \mathbf E, \mathbf D, \mathbf H)$ such that
\begin{subequations}
\label{eq:BDEH}
\begin{align}
 B^a & = \alpha \, {^*F}^{a0} \\
 E_a & = \frac{\alpha}{2} \, e_{abc} \, c \, {^*F}^{bc} \\
 D^a & = \varepsilon_0 \, c \, \alpha \, F^{a0} \\
 H_a & = - \frac{\alpha}{2\,\mu_0} \, e_{abc} \, F^{bc}
\end{align}
\end{subequations}
$\varepsilon_0$ is the vacuum permittivity and $\mu_0$ the vacuum permeability. $e_{abc} = \sqrt{\gamma} \, \varepsilon_{abc}$ is the fully antisymmetric spatial tensor and $\varepsilon_{abc}$ the three dimensional Levi-Civita symbol. The contravariant analog is $e^{abc} = \varepsilon^{abc}/\sqrt{\gamma}$. Relations eq.~(\ref{eq:BDEH}) can be inverted such that
\begin{subequations}
\begin{align}
 {^*F}^{a0} & = \frac{B^a}{\alpha} \\
 {^*F}^{ab} & = \frac{1}{c\,\alpha} \, e^{abc} \, E_c = \frac{1}{c\,\sqrt{-g}} \, \varepsilon^{abc} \, E_c \\
 F^{a0} & = \frac{D^a}{\varepsilon_0 \, c \, \alpha} \\
 F^{ab} & = - \frac{\mu_0}{\alpha} \, e^{abc} \, H_c = - \frac{\mu_0}{\sqrt{-g}} \, \varepsilon^{abc} \, H_c
\end{align}
\end{subequations}
These three dimensional vectors can be recast into
\begin{subequations}
\begin{align}
 E_a & = c \, F_{0a} \\
 H_a & = \frac{{^*F}_{0a}}{\mu_0} \\
 B^a & = - \frac{1}{2} \, e^{abc} \, F_{bc} \\
 D^a & = \frac{\varepsilon_0 \, c}{2} \, e^{abc} \, {^*F}_{bc}
\end{align}
\end{subequations}
These expressions are also easily inverted such that
\begin{subequations}
\begin{align}
  F_{0a} & = \frac{E_a}{c} \\
  {^*F}_{0a} & = \mu_0 \, H_a \\
  F_{ab} & = - e_{abc} \, B^c = - \sqrt{\gamma} \, \varepsilon_{abc} \, B^c \\
  {^*F}_{ab} & = \frac{e_{abc}}{\varepsilon_0\,c} \, D^c = \frac{\sqrt{\gamma}}{\varepsilon_0\,c} \, \varepsilon_{abc} \, D^c
\end{align}
\end{subequations}
All these antisymmetric tensors are summarized in appendix~\ref{app:metric}. With these definitions of the spatial vectors, Maxwell equations take a more traditional form in the curved three dimensional space. The system reads
\begin{subequations}
\begin{align}
\label{eq:Maxwell1}
 \divg \mathbf B & = 0 \\
\label{eq:Maxwell2}
 \rot \mathbf E & = - \frac{1}{\sqrt{\gamma}} \, \partial_t (\sqrt{\gamma} \, \mathbf B) \\
\label{eq:Maxwell3}
 \divg \mathbf D & = \rho \\
\label{eq:Maxwell4}
 \rot \mathbf H & = \mathbf J + \frac{1}{\sqrt{\gamma}} \, \partial_t (\sqrt{\gamma} \, \mathbf D)
\end{align}
\end{subequations}
The source terms $(\rho, \mathbf J)$ are given by
\begin{subequations}
\begin{align}
  \rho \, c & \equiv \alpha \, I^0 \\
  J^a & \equiv \alpha \, I^a
\end{align}
\end{subequations}
$I^k$ being the 4-current density. The above differential operators should be understood as defined in a three dimensional curved space, the absolute space with associated spatial metric~$\gamma_{ab}$, such that
\begin{subequations}
\begin{align}
 \divg \mathbf B & \equiv \frac{1}{\sqrt{\gamma}} \, \partial_a (\sqrt{\gamma} \, B^a) \\
 \rot \mathbf E & \equiv e^{abc} \, \partial_b \, E_c \\
 \mathbf E \times \mathbf B & \equiv e^{abc} \, E_b \, B_c
\end{align}
\end{subequations}
The special case of a diagonal spatial metric is given in appendix~\ref{app:operateur}. The three dimensional vector fields are not independent, they are related by two important constitutive relations, namely
\begin{subequations}
\label{eq:Constitutive}
\begin{align}
\label{eq:ConstitutiveE}
  \varepsilon_0 \, \mathbf E & = \alpha \, \mathbf D + \varepsilon_0\,c\,\mathbf\beta \times \mathbf B \\
\label{eq:ConstitutiveH}
  \mu_0 \, \mathbf H & = \alpha \, \mathbf B - \frac{\mathbf\beta \times \mathbf D}{\varepsilon_0\,c}
\end{align}
\end{subequations}
The curvature of absolute space is taken into account by the lapse function factor~$\alpha$ in the first term on the right-hand side and the frame dragging effect is included in the second term, the cross-product between the shift vector~$\mathbf\beta$ and the fields. We see that $(\mathbf D, \mathbf B)$ are the fundamental fields, actually those measured by a FIDO, see below.

\subsection{Force-free conditions}

The source terms have not yet been specified. Having in mind to apply the above equations to the pulsar magnetosphere, we give the expressions for the current in the limit of a force-free plasma, neglecting inertia and pressure. The force-free condition in covariant form reads
\begin{equation}
 F_{ik} \, I^k = 0
\end{equation}
and in the 3+1~formalism it becomes
\begin{subequations}
\begin{align}
  \mathbf J \cdot \mathbf E & = 0 \\
  \rho \, \mathbf E + \mathbf J \times \mathbf B & = \mathbf 0
\end{align}
\end{subequations}
which implies $\mathbf E \cdot \mathbf B = 0 $ and therefore also $\mathbf D \cdot \mathbf B = 0 $. As in the special relativistic case, the current density is found to be, see the derivation for instance in \cite{2011MNRAS.418L..94K}
\begin{equation}
 \mathbf J = \rho \, \frac{\mathbf E \times \mathbf B}{B^2} + \frac{\mathbf B \cdot \rot \mathbf H - \mathbf D \cdot \rot \mathbf E}{B^2} \, \mathbf B
\end{equation}
Because $c\,B^a={^*F}^{ak}\,n_k$ and $D^a/\varepsilon_0=F^{ak}\,n_k$, $\mathbf B$ and $\mathbf D/\varepsilon_0$ can be interpreted as the magnetic and electric field respectively as measured by the FIDO. Moreover
\begin{equation}
  \label{eq:DensiteFIDO}
  I^k \, n_k = \rho \, c^2  
\end{equation}
thus $\rho$ is the electric charge density as measured by this same observer. Using the projection tensor defined by
\begin{equation}
  \label{eq:Projection}
  p_i^k = \delta_i^k - \frac{n_i\,n^k}{c^2}
\end{equation}
its electric current density~$\mathbf j$ is given by
\begin{equation}
  \label{eq:CourantFIDO}
  \alpha \, \mathbf j = \mathbf J + \rho \, c \, \mathbf \beta
\end{equation}
Maxwell equations~(\ref{eq:Maxwell1})-(\ref{eq:Maxwell4}), the constitutive relations (\ref{eq:ConstitutiveE}),(\ref{eq:ConstitutiveH}) and the prescription for the source terms set the background system to be solved for any prescribed metric. In the next section, we show how to solve this system in a simple way by introducing a vector spherical harmonic basis in curved space as summarized in appendix~\ref{app:HSV}.

For the remainder of this paper, we will only focus on the vacuum field solutions, leaving the force-free case for future work. Note that we choose to keep all physical constants in the formulas because this helps to check easier the consistency with dimensionality of the equations.

\section{ELECTROMAGNETIC FIELD OF AN ALIGNED DIPOLE}
\label{sec:Aligne}

The system to be solved being linear, we treat separately the aligned and the perpendicular case, the general oblique configuration being a weighted linear superposition of both solutions. We first address the simple static and rotating aligned dipole magnetic field before investigating the interesting perpendicular rotator as a special case of an oblique rotator.

\subsection{Static dipole}

We start with a non rotating neutron star, setting the spin to zero, $a=0$, therefore $\mathbf \beta = \mathbf 0$, followed by a simplification of the constitutive relations. The electric field vanishes, thus $\mathbf E = \mathbf D = 0$ whereas $\mu_0 \, \mathbf H = \alpha \, \mathbf B$. As a consequence, the magnetic field satisfies the static ($\partial_t=0$) Maxwell equations given by
\begin{subequations}
\begin{align}
  \label{eq:Static1}
  \divg \mathbf B & = 0 \\
  \label{eq:Static2}
  \rot (\alpha \, \mathbf B ) & = 0
\end{align}
\end{subequations}
Far from the neutron star, we expect to retrieve the flat space-time expression for the dipole magnetic field with magnetic moment $\mathbf \mu$ or, written explicitly,
\begin{equation}
\label{eq:BdipolePlat}
 \mathbf B = \frac{\mu_0}{4\,\pi\,r^3} \, \left[ \frac{3\,(\mathbf \mu \cdot \mathbf r) \, \mathbf r}{r^2} - \mathbf \mu \right] = - \frac{\mu_0\,\mu}{4\,\pi} \, \sqrt{\frac{8\,\pi}{3}} \, \mathrm{Re} \left[ \rot \frac{\mathbf \Phi_{1,0}}{r^2}\right] 
\end{equation}
In curved space-time, the meaning of a dipole field needs to be explicitly defined. We take as a definition for the dipolar magnetic field the one which is expressed only with the first vector spherical harmonic $\mathbf \Phi_{1,0}$ corresponding to the mode $(l,m)=(1,0)$ according to its flat space-time expression Eq.~(\ref{eq:BdipolePlat}). This is valid for a symmetry around the $z$-axis because $m=0$. The perpendicular case or more generally the oblique rotator would include the mode $(l,m)=(1,1)$ for the dipolar field. This will be done in section~\ref{sec:Orthogonal}.
Thus we expand the magnetic field according to the divergencelessness prescription and look for a separable solution with the prescription
\begin{equation}
  \label{eq:Bstatic1}
  \mathbf B = \mathrm{Re} \left[ \rot ( f_{1,0}^B(r) \, \mathbf \Phi_{1,0} ) \right] 
\end{equation}
with the boundary condition
\begin{equation}
  \label{eq:Bstatic2}
  \lim\limits_{r\to+\infty} f_{1,0}^B(r) = - \frac{\mu_0\,\mu}{4\,\pi\,r^2} \, \sqrt{\frac{8\,\pi}{3}}
\end{equation}
$\mathbf \Phi_{1,0}$ is a vector spherical harmonic, see for instance \cite{2012MNRAS.424..605P}. The vector spherical harmonics being proper functions of the curl linear differential operator insure that such separable solutions do indeed exist. These linear algebra properties are absolutely fundamental and make vector spherical harmonics extremely useful to solve linear partial differential equations involving vector fields. Note that $f_{1,0}^B(r)$ is the unique unknown in this simple problem and depends only on the radial coordinate~$r$. Eq.~(\ref{eq:Static1}) is automatically satisfied by construction whereas inserting the expansion eq.~(\ref{eq:Bstatic1}) into eq.~(\ref{eq:Static2}) following the property eq.~(\ref{eq:RotRotPhi}) of appendix~\ref{app:HSV} (for $l=1$) leads to a second order linear ordinary differential equation for the scalar function~$f_{1,0}^B$ such that
\begin{equation}
  \label{eq:Laplacef10}
  \partial_r(\alpha^2\,\partial_r(r\,f_{1,0}^B)) - \frac{2}{r} \, f_{1,0}^B = 0
\end{equation}
The exact solution to this boundary problem which asymptotes to the flat dipole at large distances as prescribed by eq.~(\ref{eq:Bstatic2}) is given by
\begin{equation}
  \label{eq:DipoleSchwarzf10}
  f_{1,0}^{B({\rm dip})} = \frac{\mu_0\,\mu}{4\,\pi} \, \sqrt{\frac{8\,\pi}{3}} \, \frac{3\,r}{R_s^3} \, \left[ {\rm ln} \left( 1 - \frac{R_s}{r} \right) + \frac{R_s}{r} + \frac{R_s^2}{2\,r^2}\right]
\end{equation}
which corresponds to the solution shown in~\cite{1964ZETF...47..1030G}. The non vanishing magnetic field components are
\begin{subequations}
\label{eq:MagneticStatic}
\begin{align}
  \label{eq:MagneticStaticR}
  B^{\hat r} & = - 6 \, \frac{\mu_0}{4\,\pi} \, \left[ {\rm ln} \left( 1 - \frac{R_s}{r} \right) + \frac{R_s}{r} + \frac{R_s^2}{2\,r^2} \right] \, \frac{\mu\,\cos\vartheta}{R_s^3} \\
  \label{eq:MagneticStaticT}
  B^{\hat \vartheta} & = 3 \, \frac{\mu_0}{4\,\pi} \, \left[ 2 \, \sqrt{ 1 - \frac{R_s}{r}} \, {\rm ln} \left( 1 - \frac{R_s}{r} \right) + \frac{R_s}{r} \, \frac{2\,r-R_s}{\sqrt{r\,(r-R_s)}} \right] \, \frac{\mu\,\sin\vartheta}{R_s^3}
\end{align}
\end{subequations}
Corrections to first order compared to flat space-time are
\begin{subequations}
\begin{align}
  \label{eq:Correction}
    B^{\hat r} & = \frac{\mu_0}{4\,\pi} \, \frac{2\,\mu\,\cos\vartheta}{r^3} \, \left[ 1 + \frac{3}{4} \, \frac{R_s}{r} + o\left( \frac{R_s}{r} \right) \right] \\
  B^{\hat \vartheta} & = \frac{\mu_0}{4\,\pi} \, \frac{\mu\,\sin\vartheta}{r^3} \, \left[ 1 + \frac{R_s}{r} + o\left( \frac{R_s}{r} \right) \right]
\end{align}
\end{subequations}
This first example shows how easy it is to compute the solution once the expansion onto vector spherical harmonics has been performed and knowing their properties and action on linear differential operators.

\subsection{Rotating dipole}

Next we consider the more useful case of a rotating magnetic dipole with magnetic moment aligned to the rotation axis. Now the situation becomes much more involved. First, rotation induces an electric field and secondly frame dragging effects mix electric and magnetic fields through the constitutive relations eq.~(\ref{eq:Constitutive}). To demonstrate how our formalism works, we decided to split the task in two steps. First we neglect frame dragging effects and look solely for the induced electric field. In a second stage, we add frame dragging.

\subsubsection{A pedestrian way}

Frame dragging effects could become important and should be included. Nevertheless, before dealing with the most general expression including frame dragging, we think it is educational to introduce the reasoning by hand and work out a low order expansion explicitly without any frame dragging effect. This would be acceptable for sufficiently low rotation and we can in the first stage neglect the shift vector setting $\mathbf\beta=0$ as in the previous paragraph. Maxwell equations then become
\begin{subequations}
\begin{align}
  \label{eq:Tournant1}
  \divg \mathbf D & = 0 \\
  \label{eq:Tournant2}
  \rot (\alpha \, \mathbf D ) & = 0 \\
  \divg \mathbf B & = 0 \\
  \rot (\alpha \, \mathbf B ) & = 0
\end{align}
\end{subequations}
These equations are particularly straightforward to solve because it represents a decoupled system of two unknown vector fields, one for $\mathbf D$ and one for $\mathbf B$. From the flat space-time solution, we know that the electric field will be quadrupolar which means only one mode is present namely $(l,m)=(2,0)$ for the axisymmetric case. Thus we expand both fields according to
\begin{subequations}
\begin{align}
  \label{eq:TournantD2}
  \mathbf D & = \mathrm{Re} \left[ \rot ( f_{2,0}^D \, \mathbf \Phi_{2,0} ) \right] \\
  \label{eq:TournantB1}
  \mathbf B & = \mathrm{Re} \left[ \rot ( f_{1,0}^B \, \mathbf \Phi_{1,0} ) \right] 
\end{align}
\end{subequations}
This expansion insure automatically and analytically the divergencelessness nature of both $\mathbf D$ and $\mathbf B$. Moreover, these expressions lead as in the previous static regime to a separable solution for both the electric and magnetic field. Straightforward calculations show that $f_{1,0}^B$ again satisfies eq.~(\ref{eq:Laplacef10}) whereas $f_{2,0}^D$ has to be solution of another second order linear differential equation given by
\begin{equation}
  \label{eq:Laplacef20}
  \partial_r(\alpha^2\,\partial_r(r\,f_{2,0}^D)) - \frac{6}{r} \, f_{2,0}^D = 0
\end{equation}
It is obtained by inserting the expansion eq.~(\ref{eq:TournantD2}) into eq.~(\ref{eq:Tournant2}) following the property eq.~(\ref{eq:RotRotPhi}) of appendix~\ref{app:HSV} but now for $l=2$. The exact solution of this homogeneous linear differential equation and vanishing at infinity reads
\begin{equation}
  \label{eq:DipoleSchwarzf20}
  f_{2,0}^D = \frac{K}{R_s^2\,r} \, \left[ 6 \, \frac{r^2}{R_s^2} \, \left( 3 - 4 \, \frac{r}{R_s} \right) \, {\rm ln} \left( 1 - \frac{R_s}{r} \right) + 1 + 6 \, \frac{r}{R_s} \, \left( 1 - 4 \, \frac{r}{R_s} \right) \right]
\end{equation}
where $K$ is a constant to be determined from the boundary conditions at the surface of the neutron star. We now discuss this inner boundary condition in more details. Inside a perfectly conducting star, the rotation of the plasma induces an electric field $\mathbf E$ which satisfies
\begin{equation}
  \label{eq:CLE}
  \mathbf E + r\,\Omega\,\sin\vartheta \,\ephi \times \mathbf B = 0
\end{equation}
This implies an electric field as measured by a FIDO given by
\begin{equation}
  \label{eq:CLD}
  \mathbf D = - \varepsilon_0 \, \frac{\tilde{\omega}}{\alpha}\,r\,\sin\vartheta \,\ephi \times \mathbf B = \varepsilon_0 \, c \, \frac{\tilde{\omega}}{\alpha} \, \frac{\mathbf \beta}{\omega} \times \mathbf B
\end{equation}
For this FIDO, the electromagnetic field symbolized by $(\mathbf D, \mathbf B)$ has to verify the jump conditions across an interface as in flat space-time. In other words, the magnetic field component normal to the surface and the electric field components lying in the plane of the interface are continuous functions. More explicitly, the normal component~$B^{\hat r}$ and the tangential components~$(D^{\hat \vartheta}, D^{\hat \varphi})$ have to be continuous across the stellar surface. By construction, it can be verified by projection of eq.~(\ref{eq:TournantD2}) onto $\ephi$ that the component $D^{\hat \varphi}$ remains zero in the exterior vacuum space, as it is inside the star. Note that this remark is consistent with the projection of eq.~(\ref{eq:CLD}) onto $\ephi$. For the other tangential component, by projection of eq.~(\ref{eq:CLD}) onto $\etheta$ we have to enforce the condition
\begin{equation}
  \label{eq:LimiteFD20}
  D^{\hat \vartheta} = - \varepsilon_0 \, \frac{\tilde{\omega}}{\alpha}\,r\,\sin\vartheta \, B^{\hat r}
\end{equation}
This has to be compared with the projection of eq.~(\ref{eq:TournantD2}) onto $\etheta$ and given by
\begin{equation}
  \label{eq:LimiteD20}
  D^{\hat \vartheta} = \frac{3}{2} \, \sqrt{\frac{5}{6\,\pi}} \, \frac{\alpha}{r} \, \partial_r(r\,f^D_{2,0}) \, \sin\vartheta \, \cos \vartheta
\end{equation}
In order to deduce the constant of integration~$K$ in eq.~(\ref{eq:DipoleSchwarzf20}), eq.~(\ref{eq:LimiteFD20}) and~(\ref{eq:LimiteD20}) should be compared at the stellar surface setting $r=R$. $B^{\hat r}$ is known from the static dipole solution and given by eq.~(\ref{eq:MagneticStaticR}). By direct calculation from eq.~(\ref{eq:DipoleSchwarzf20}) we arrive at
\begin{equation}
 \left.\partial_r(r\,f^D_{2,0})\right|_{r=R} = 36 \, \frac{K\,R}{\Rs^4} \, \left[ \left( 1 - 2\,\frac{R}{R_s} \right) \, \ln\alpha_R^2 - 2 - \frac{R_s^2}{6\,R^2\,\alpha_R^2} \right]
\end{equation}
The constant $K$ then follows immediately from the above condition. We get
\begin{equation}
  \label{eq:K}
  K = \frac{\varepsilon_0 \, \mu_0 \, \mu}{4\,\pi} \, \frac{1}{9} \, \sqrt{\frac{6\,\pi}{5}} \, R_s \, R \,
  \frac{\tilde{\omega}_R}{\alpha_R^2} \, C_1 \, C_2
\end{equation}
where 
\begin{subequations}
\begin{align}
  \label{eq:C1}
  \alpha_R & = \sqrt{ 1 - \frac{R_s}{R} }\\
  \omega_R & = \frac{a\,R_s\,c}{R^3} \\
  \tilde{\omega}_R & = \Omega - \omega_R \\
  C_1 & = \ln\alpha_R^2 + \frac{R_s}{R} + \frac{R_s^2}{2\,R^2} \\
  C_2 & = \left[ \left( 1 - 2\,\frac{R}{R_s} \right) \, \ln\alpha_R^2 - 2 -
  \frac{R_s^2}{6\,R^2\,\alpha_R^2} \right]^{-1}
\end{align}
\end{subequations}
The magnetic field remains the same as for the static dipole and the electric field yields
\begin{subequations}
\begin{align}
  \label{eq:Drot1}
  D^{\hat r} & = - \frac{\varepsilon_0 \, \mu_0 \, \mu}{4\,\pi} \, \frac{R}{R_s^3} \,
  \frac{\tilde{\omega}_R}{\alpha_R^2} \, C_1 \, C_2 \, \left[ \left( 3
      - 4\,\frac{r}{R_s} \right) \, \ln\alpha^2 + \frac{R_s^2}{6\,r^2}
    + \frac{R_s}{r} - 4 \right] \, ( 3\,\cos^2\vartheta - 1 ) \\
  D^{\hat \vartheta} & = 6 \, \frac{\varepsilon_0 \, \mu_0 \, \mu}{4\,\pi} \, \frac{R}{R_s^3} \,
  \frac{\tilde{\omega}_R}{\alpha_R^2} \, \alpha \, C_1 \, C_2 \, \left[
    \left( 1 - 2\,\frac{r}{R_s} \right) \, \ln\alpha^2 - 2 -
    \frac{R_s^2}{6\,r^2\,\alpha^2} \right] \, \cos\vartheta \, \sin\vartheta \\
  D^{\hat \varphi} & = 0
\end{align}
\end{subequations}
So far, we did not include any frame dragging effect symbolized by the cross product in the constitutive relations eq.~(\ref{eq:ConstitutiveE}) and~(\ref{eq:ConstitutiveH}). Now we proceed to the inclusion of the frame dragging effect to look for more accurate solutions taking explicitly into account the rotation of the neutron star. Because these constitutive relations and the vacuum Maxwell equations are linear, we use a power series expansion of the unknown vector fields with respect to a small adimensionalized parameter~$\varepsilon$ which is related to the neutron star spin such that $\varepsilon=O(\Omega)$. Any vector field $\mathbf{V}$ is expanded into
\begin{equation}
  \label{eq:Seri}
  \mathbf{V} = \sum_{k\ge0} \varepsilon^k \, \mathbf{V}_k = \mathbf{V}_0 + \sum_{k\ge1} \varepsilon^k \, \mathbf{V}_k
\end{equation}
$\mathbf{V}_0$ is the static field for the non rotating star. Thus, to zero-th order, the electric field vanishes, $\mathbf{D}_0 = \mathbf{E}_0 = 0$. They are at least first order in $\Omega$. The shift vector is a quantity of first order so we write it as $\mathbf{\beta} = \varepsilon \, \mathbf{\beta}_1$. From the constitutive relations, we get the $k$-th order of the auxiliary electric field for $k\ge1$ as
\begin{equation}
  \label{eq:Ek1}
  \varepsilon_0 \, \mathbf{E}_k = \alpha \, \mathbf{D}_k + \varepsilon_0 \, c \, \mathbf{\beta}_1 \times \mathbf{B}_{k-1}
\end{equation}
and for the $k$-th order of the auxiliary magnetic field
\begin{equation}
  \label{eq:Ek2}
  \mu_0 \, \mathbf{H}_k = \alpha \, \mathbf{B}_k - \frac{\mathbf{\beta}_1 \times \mathbf{D}_{k-1}}{\varepsilon_0 \, c}
\end{equation}
Moreover, for any order, we have the constraints
\begin{subequations}
\begin{align}
  \label{eq:Contraintek}
  \divg \mathbf{B}_k & = \divg \mathbf{D}_k = 0 \\
  \rot \mathbf{H}_k & = \rot \mathbf{E}_k = 0
\end{align}
\end{subequations}
As a consequence, we obtain a hierarchical set of partial differential equations for the fields~$\{\mathbf{B}_k, \mathbf{D}_k\}$ such that
\begin{subequations}
\begin{align}
  \label{eq:hierarchy1}
  \rot (\alpha \, \mathbf{D}_k) & = - \varepsilon_0 \, c \, \rot (\mathbf{\beta}_1 \times \mathbf{B}_{k-1}) \\
  \label{eq:hierarchy2}
  \rot (\alpha \, \mathbf{B}_k) & = \frac{1}{\varepsilon_0 \, c} \, \rot (\mathbf{\beta}_1 \times \mathbf{D}_{k-1})
\end{align}
\end{subequations}
for $k\ge1$. The initialisation for $k=0$ corresponds to the static dipole with $\mathbf B_0$ given by eq.~(\ref{eq:DipoleSchwarzf20}), therefore $\mathbf{D}_0 = \mathbf{E}_0 = 0$ as expected. We immediately conclude that the first perturbation in magnetic field corresponds to a second order term symbolized by~$\mathbf{B}_2$ ($\mathbf{B}_1=0$). We look for the first order perturbation in electric field corresponding to an electric quadrupole with $(l,m)=(2,0)$ such that
\begin{equation}
  \label{eq:Tournant}
  \mathbf{D}_1 = \rot ( f_{2,0}^D \, \mathbf \Phi_{2,0} ) 
\end{equation}
From now on, we suppress the real part symbol, it should be understood that the physical fields correspond to the real parts of the expressions derived below. Inserting this expansion into eq.~(\ref{eq:hierarchy1}) with $k=1$, the function~$f_{2,0}^D$ is  solution of the following second order inhomogeneous linear ordinary differential equation
\begin{equation}
  \label{eq:LaplaceSourcef20}
  \partial_r(\alpha^2\,\partial_r(r\,f_{2,0}^D)) - \frac{6}{r} \,
  f_{2,0}^D = 12 \, \frac{\varepsilon_0 \, \mu_0 \, \mu}{4\,\pi} \, \sqrt{\frac{6\,\pi}{5}} \,  \frac{a\,c}{R_s^2\,r^2} \, \left[ \ln\alpha^2 + \frac{R_s}{r} +
    \frac{R_s^2}{2\,r^2} \right]
\end{equation}
The right hand side is obtained from the property eq.~(\ref{eq:RotBETARotVHS}). To solve this equation, we use standard techniques. First we look for the general solution to the homogeneous equation which is nothing else than eq.~(\ref{eq:Laplacef20}) with the subsequent solution eq.~(\ref{eq:DipoleSchwarzf20}), which we write here as $f_{2,0}^{D(h)}$. Next a peculiar solution of the inhomogeneous eq.~(\ref{eq:LaplaceSourcef20}) and vanishing at infinity is given by
\begin{equation}
  \label{eq:SolPart20}
  f_{2,0}^{D(p)} = - 2 \, \frac{\varepsilon_0 \, \mu_0 \, \mu}{4\,\pi} \, \sqrt{\frac{6\,\pi}{5}} \, \frac{a\,c}{R_s^2\,r} \, \left[ \ln\alpha^2 + \frac{R_s}{r} \right]
\end{equation}
In order to satisfy the boundary condition on the star, we also need the following expression
\begin{equation}
 \partial_r (r\,f_{2,0}^{D(p)}) = - 2 \, \frac{\varepsilon_0 \, \mu_0 \, \mu}{4\,\pi} \, \sqrt{\frac{6\,\pi}{5}} \, \frac{a\,c}{R^3 \, \alpha_R^2}
\end{equation}
in order to compute $\partial_r (r\,f_{2,0}^{D})$ in eq.~(\ref{eq:LimiteD20}) from $f_{2,0}^{D} = f_{2,0}^{D(h)} + f_{2,0}^{D(p)}$. The constant~$K$ will be determined from the inner boundary condition, now taking the frame dragging effect into account because of the presence of the peculiar solution~$f_{2,0}^{D(p)}$, it has to be set to
\begin{equation}
  \label{eq:Kb1}
  K = \frac{\varepsilon_0 \, \mu_0 \, \mu}{4\,\pi} \, \frac{C_2}{9\,\alpha_R^2} \, \sqrt{\frac{6\,\pi}{5}} \, \left[ R_s \, R \, \tilde{\omega}_R \, C_1 + \frac{1}{2}\, \, \frac{\omega_R\,R_s^3}{R} \right]
\end{equation}
Putting all pieces together, the full solution $f_{2,0}^{D} = f_{2,0}^{D(h)} + f_{2,0}^{D(p)}$ reads
\begin{multline}
 f_{2,0}^D = \frac{K}{R_s^2\,r} \, \left[ 6 \, \frac{r^2}{R_s^2} \, \left( 3 - 4 \, \frac{r}{R_s} \right) \, {\rm ln} \left( 1 - \frac{R_s}{r} \right) + 1 + 6 \, \frac{r}{R_s} \, \left( 1 - 4 \, \frac{r}{R_s} \right) \right] \\
 - 2 \, \frac{\varepsilon_0 \, \mu_0 \, \mu}{4\,\pi} \, \sqrt{\frac{6\,\pi}{5}} \, \frac{a\,c}{R_s^2\,r} \, \left[ \ln\alpha^2 + \frac{R_s}{r} \right]
\end{multline}
Taking the value of the constant~$K$ into account, we finally get
\begin{multline}
   f_{2,0}^{D({\rm quad})} = \frac{\varepsilon_0 \, \mu_0 \, \mu}{4\,\pi\,r} \, \sqrt{\frac{6\,\pi}{5}} \left\{ \frac{C_2}{18\,\alpha_R^2} \, \left( \frac{\omega_R\,R_s}{R} + 2 \, C_1 \, \frac{\tilde{\omega_R}\,R}{R_s} \right) \right. \times \\
   \times \left. \left[ 6 \, \frac{r^2}{R_s^2} \, \left( 3 - 4 \, \frac{r}{R_s} \right) \, {\rm ln} \left( 1 - \frac{R_s}{r} \right) + 1 + 6 \, \frac{r}{R_s} \, \left( 1 - 4 \, \frac{r}{R_s} \right) \right] - 2 \, \frac{\omega\,r^3}{R_s^3} \, \left( \ln\alpha^2 + \frac{R_s}{r} \right) \right\}
\end{multline}
If we separate the frame dragging effect~$\omega$ from the pure rotation~$\Omega$, we get
\begin{multline}
   f_{2,0}^{D({\rm quad})} = \frac{\varepsilon_0 \, \mu_0 \, \mu}{4\,\pi} \, \sqrt{\frac{6\,\pi}{5}} \left\{ \frac{2}{3} \, \frac{C_1\,C_2\,\Omega\,R\,r}{\alpha_R^2\,R_s^3} \,  \left[ \left( 3
      - 4\,\frac{r}{R_s} \right) \, \ln\alpha^2 + \frac{R_s^2}{6\,r^2} + \frac{R_s}{r} - 4 \right] - \right. \\ 
\left. 2 \, \frac{\omega\,r^4}{R_s^5} \, \left( \frac{R_s^2}{r^2} \, \left( \ln\alpha^2 + \frac{R_s}{r} \right) + \frac{C_2\,R_s^2}{3\,\alpha_R^2\,R^2} \, \left( \ln\alpha_R^2 + \frac{R_s}{R} \right) \, \times \right. \right. \\ 
\left.\left.\left[ \left( 3 - 4\,\frac{r}{R_s} \right) \, \ln\alpha^2 + \frac{R_s^2}{6\,r^2} + \frac{R_s}{r} - 4 \right]  \right) \right\}
\end{multline}
The components of the electric field are then
\begin{subequations}
\begin{align}
  \label{eq:Drot2}
  D^{\hat r} & =  - \sqrt{\frac{5}{4\,\pi}} \, \frac{\sqrt{6}}{r} \, f_{2,0}^{D({\rm quad})} \, P_2(\cos\vartheta) \\
  D^{\hat \vartheta} & = \frac{3}{2} \, \sqrt{\frac{5}{6\,\pi}} \, \frac{\alpha}{r} \, \partial_r(r\, f_{2,0}^{D({\rm quad})}) \, \cos\vartheta \, \sin\vartheta \\ 
  D^{\hat \varphi} & = 0 
\end{align}
with the radial derivative given explicitly by
\begin{align}
\partial_r(r\, f_{2,0}^{D({\rm quad})}) & = \frac{2}{3} \, \sqrt{\frac{6\,\pi}{5}} \, \frac{\varepsilon_0 \, \mu_0 \, \mu \, r}{4\,\pi} \, \left\{ \frac{6\,C_1\,C_2\,\Omega\,R}{\alpha_R^2\,R_s^3} \, \left[ \left( 1 - 2 \, \frac{r}{R_s} \right) \, {\rm ln} \alpha^2 - 2 - \frac{R_s^2}{6\,\alpha^2\,r^2} \right] \right. \\
  & \left. - \frac{\omega\,r^3}{R_s^5} \left( \frac{6\,C_2\,R_s^2}{\alpha_R^2\,R^2} \, \left( \ln\alpha_R^2 + \frac{R_s}{R} \right) \, \left[ \left( 1 - 2 \, \frac{r}{R_s} \right) \, {\rm ln} \alpha^2 - 2 - \frac{R_s^2}{6\,\alpha^2\,r^2} \right] + 3 \, \frac{R_s^4}{\alpha^2\,r^4} \right) \right\} \nonumber
\end{align}
\end{subequations}
These expressions are the same as equations (124)-(125)-(126) in \cite{2001MNRAS.322..723R} specialized to the aligned rotator. In the newtonian limit we find
\begin{subequations}
\begin{align}
  \label{eq:DrotNewt}
  D^{\hat r} & = - \frac{\Omega \, B \, R^5}{r^4} \, ( 3 \, \cos^2 \vartheta - 1 ) \\
  D^{\hat \vartheta} & = - \frac{\Omega \, B \, R^5}{r^4} \, 2 \, \cos \vartheta \, \sin \vartheta \\
  D^{\hat \varphi} & = 0
\end{align}
\end{subequations}
as it should be.

\subsubsection{General formalism}

The properties of the vector spherical harmonics in curved space allow us to derive in a systematic way the relations between the expansion coefficients of~$\{\mathbf{B}_k, \mathbf{D}_k\}$. Because of the axisymmetry of the problem, there are no toroidal components of neither the magnetic nor the electric part. Therefore, all coefficients with $m>0$ vanish. Thus we expand both fields according to
\begin{subequations}
\begin{align}
 \mathbf{D}_k & = \sum_{l\geq1} \rot ( f_{l,0}^{D(k)} \, \mathbf \Phi_{l,0} ) \\
 \mathbf{B}_k & = \sum_{l\geq1} \rot ( f_{l,0}^{B(k)} \, \mathbf \Phi_{l,0} )
\end{align}
\end{subequations}
the superscript~$(k)$ denotes the order of the expansion in the spin parameter, related to the frame dragging effect. Injecting those expressions into eqs.~(\ref{eq:hierarchy1}) and (\ref{eq:hierarchy2}), we get for $(l,k)\geq1$ according to eq.~(\ref{eq:RotBETARotVHS}) in appendix~\ref{app:HSV}
\begin{subequations}
\begin{multline}
\label{eq:fl0Dk}
 \partial_r(\alpha^2\,\partial_r(r\,f_{l,0}^{D(k)})) - \frac{l(l+1)}{r} \, f_{l,0}^{D(k)} = \\
 3 \, \varepsilon_0 \, \omega \, \left[ l \sqrt{\frac{(l-1)(l+1)}{(2l-1)(2l+1)}} \, f_{l-1,0}^{B(k-1)} - (l+1) \sqrt{\frac{l(l+2)}{(2l+3)(2l+1)}} \, f_{l+1,0}^{B(k-1)} \right]
\end{multline}
\begin{multline}
 \label{eq:fl0Bk}
 \partial_r(\alpha^2\,\partial_r(r\,f_{l,0}^{B(k)})) - \frac{l(l+1)}{r} \, f_{l,0}^{B(k)} = \\
 - 3 \, \mu_0 \, \omega \, \left[ l \sqrt{\frac{(l-1)(l+1)}{(2l-1)(2l+1)}} \, f_{l-1,0}^{D(k-1)}- (l+1) \sqrt{\frac{l(l+2)}{(2l+3)(2l+1)}} \, f_{l+1,0}^{D(k-1)} \right]
\end{multline} 
\end{subequations}
It is understood that $f_{0,0}^{D(k)} = f_{0,0}^{B(k)} = 0$. The very important fact about this hierarchical system of second order linear partial differential equations relating the $f_{l,0}^{D(k)}$ to the $f_{l,0}^{D(k)}$ is its \textit{uncoupled} nature. Indeed the coefficients $f_{l,0}^{D(k)}$ and $f_{l,0}^{B(k)}$ are related to the immediately lowest order expansion coefficients $f_{l,0}^{D(k-1)}$ and $f_{l,0}^{B(k-1)}$. Consequently, we can find the solution to any order by simply computing more coefficients. The recurrence starts with the static aligned magnetic dipole which is the zero-th order approximation of the solution, with subscript~$(0)$. As already noted in the previous paragraph, the electric field is a first order effect at least.

For concreteness, let us work out the approximate solution to third order, i.e. including to vector spherical harmonic functions in the expansion of both the electric and the magnetic field. At first glance, this seems a rather high degree of accuracy for such solution in contrast to the first order expansion of the background metric. Nevertheless, we have in mind to use such results as a benchmark to test forthcoming general relativistic electromagnetic solvers in free space and in the force-free approximation in order to extend the code presented in \cite{2012MNRAS.424..605P}. This justifies our wish to reach a high degree of accuracy for the numerical solutions even if the metric is only first order accurate in the spin parameter~$a$.

The electric field is a consequence of the rotation of the star, thus to zero-th order, there is only a magnetic field, i.e. $f_{l,0}^{D(0)} = 0 $ and $f_{1,0}^{B(0)} = f_{1,0}^{B({\rm dip})}$, the dipole in Schwarzschild space-time given by eq.~(\ref{eq:DipoleSchwarzf10}), all other $f_{l,0}^{B(0)}$ with $l\geq2$ being equal to zero. The initialisation with $f_{l,0}^{D(0)} = 0 $ implies that there are no first order corrections to the magnetic field because eq.~(\ref{eq:fl0Bk}) has a vanishing right hand side. It is a linear homogeneous second order partial differential equation with zero boundary conditions on the star and at infinity. Therefore the solution vanishes identically leading to $f_{l,0}^{B(1)}=0$. The first correction comes from the coefficients $f_{l,0}^{D(1)}$ which have to satisfy eq.~(\ref{eq:fl0Dk}). There is only one inhomogeneous equation corresponding to $l=2$ with the right hand side including $f_{1,0}^{B(0)}$. If written explicitly, we retrieve eq.~(\ref{eq:LaplaceSourcef20}) namely
\begin{equation}
  \partial_r(\alpha^2\,\partial_r(r\,f_{2,0}^{D(1)})) - \frac{6}{r} \,
  f_{2,0}^{D(1)} = \frac{6}{\sqrt{5}} \, \varepsilon_0 \, \omega \, f_{1,0}^{B(0)}
\end{equation} 
with its subsequent solution. The next order includes a perturbation in the magnetic field. Indeed, the coefficients~$f_{l,0}^{B(2)}$ have to satisfy eq.~(\ref{eq:fl0Bk}) with source terms emanating only from $f_{2,0}^{D(1)}$, supplemented with vanishing boundary conditions. The two equations are inhomogeneous, namely
\begin{subequations}
\begin{align}
\label{eq:f10B2}
 \partial_r(\alpha^2\,\partial_r(r\,f_{1,0}^{B(2)})) - \frac{2}{r} \, f_{1,0}^{B(2)} & = \frac{6}{\sqrt{5}} \, \frac{\omega}{\varepsilon_0 \, c^2} \, f_{2,0}^{D(1)} \\
 \label{eq:f30B2}
 \partial_r(\alpha^2\,\partial_r(r\,f_{3,0}^{B(2)})) - \frac{12}{r} \, f_{3,0}^{B(2)} & = - 18 \, \sqrt{\frac{2}{35}} \, \frac{\omega}{\varepsilon_0 \, c^2} \, f_{2,0}^{D(1)}
\end{align}
\end{subequations}
Finally, this perturbed magnetic field will feed back to the electric field to third order with the non-vanishing coefficients satisfying
\begin{subequations}
\begin{align}
\label{eq:f20D3}
 \partial_r(\alpha^2\,\partial_r(r\,f_{2,0}^{D(3)})) - \frac{6}{r} \, f_{2,0}^{D(3)} & = \frac{6}{\sqrt{5}} \, \varepsilon_0 \, \omega \, \left[ f_{1,0}^{B(2)} - 3 \sqrt{\frac{2}{7}} \, f_{3,0}^{B(2)} \right] \\
\label{eq:f40D3}
 \partial_r(\alpha^2\,\partial_r(r\,f_{4,0}^{D(3)})) - \frac{20}{r} \, f_{4,0}^{D(3)} & = 4 \, \sqrt{\frac{15}{7}} \, \varepsilon_0 \, \omega \, f_{3,0}^{B(2)}
\end{align}
\end{subequations}
but with boundary conditions at the stellar surface according to eq.~(\ref{eq:CLD}). Equations (\ref{eq:f10B2})-(\ref{eq:f20D3}) show the hierarchical set we are led to in order to improve the solution step by step by including an increasing number of multipoles of order~$l$ in accordance with the degree of approximation desired in the spin parameter. Some of these equations can be solved analytically with source terms, but we were unable to write down simple expressions for the solution with appropriate boundary conditions except for the very few first coefficients.

Finding closed expression is a cumbersome task. Eventually, we decided to solve the above set of equations numerically by spectral methods. We expand the solutions into rational Chebyshev functions as defined in \cite{Boyd2001}. See below for the details. Our starting point is to use a finite number of multipolar coefficients in the expansion of both the fields, $N_D$ terms for $\mathbf{D}$ and $N_B$ terms for $\mathbf{B}$, writing
\begin{subequations}
\label{eq:DBDvlptAlign}
\begin{align}
 \mathbf{D} & = \sum_{l=1}^{N_D} \rot ( f_{l,0}^{D} \, \mathbf \Phi_{l,0} ) \\
 \mathbf{B} & = \sum_{l=1}^{N_B} \rot ( f_{l,0}^{B} \, \mathbf \Phi_{l,0} )
\end{align}
\end{subequations}
The order in the spin parameter, previously labelled as~$(k)$, has disappeared in the numerical solution, we do not perturb anymore according to~$a$. Each of the coefficient $f_{l,0}^{D}$ and $f_{l,0}^{B}$ has to satisfy the differential equation which is given for the magnetic field by
\begin{multline}
 \label{eq:fl0B}
 \partial_r(\alpha^2\,\partial_r(r\,f_{l,0}^{B})) - \frac{l(l+1)}{r} \, f_{l,0}^{B} = \\
 - 3 \, \frac{\omega}{\varepsilon_0 \, c^2} \, \left[ l \sqrt{\frac{(l-1)(l+1)}{(2l-1)(2l+1)}} \, f_{l-1,0}^{D}- (l+1) \sqrt{\frac{l(l+2)}{(2l+3)(2l+1)}} \, f_{l+1,0}^{D} \right]
\end{multline}
and for the electric field by
\begin{multline}
\label{eq:fl0D}
 \partial_r(\alpha^2\,\partial_r(r\,f_{l,0}^{D})) - \frac{l(l+1)}{r} \, f_{l,0}^{D} = \\
 3 \, \varepsilon_0 \, \omega \, \left[ l \sqrt{\frac{(l-1)(l+1)}{(2l-1)(2l+1)}} \, f_{l-1,0}^{B} - (l+1) \sqrt{\frac{l(l+2)}{(2l+3)(2l+1)}} \, f_{l+1,0}^{B} \right]
\end{multline}
The boundary conditions at infinity enforce vanishing coefficients whereas on the neutron star surface, we have to impose continuity of the tangential $\mathbf D$ and normal $\mathbf B$ components on the stellar surface. Introducing the expansions eq.~(\ref{eq:DBDvlptAlign}) into eq.~(\ref{eq:CLD}), then projecting along~$\etheta$ and using the useful identities for frame-dragging presented in appendix~\ref{sec:framedragging} we get the relation between the coefficients of $\mathbf D$ and $\mathbf B$ as
\begin{multline}
 \alpha^2 \, \left[ \sqrt{\frac{l+2}{l+1}} \,J_{l+1,0} \, \partial_r(r\,f_{l+1,0}^{D}) - \sqrt{\frac{l-1}{l}} \,J_{l,0} \, \partial_r(r\,f_{l-1,0}^{D}) \right] = \\
 \varepsilon_0 \, r \, \tilde{\omega} \, \left[ \sqrt{l\,(l+1)} \, ( 1 - J_{l,0}^2 - J_{l+1,0}^2 ) \, f_{l,0}^{B} - \right. \\
 \left. \sqrt{(l-2)\,(l-1)} \, J_{l,0} \, J_{l-1,0} \, f_{l-2,0}^{B} - \sqrt{(l+2)\,(l+3)} \, J_{l+1,0} \, J_{l+2,0} \, f_{l+2,0}^{B}\right] 
\end{multline}
where quantities have to be evaluated on the neutron star surface, at $r=R$. Some recurrence formulas are very useful to deal with spherical harmonics. The three recurrences used to impose the boundary conditions are
\begin{subequations}
 \begin{align}
  \sin\vartheta \, \partial_\vartheta Y_{l,m} & = l \, J_{l+1,m} \, Y_{l+1,m} - (l+1) \, J_{l,m} \, Y_{l-1,m} \\
  \cos\vartheta \, Y_{l,m} & = J_{l+1,m} \, Y_{l+1,m} + J_{l,m} \, Y_{l-1,m} \\
  \cos^2\vartheta \, Y_{l,m} & = J_{l+1,m} \, J_{l+2,m} \, Y_{l+2,m} + ( J_{l+1,m}^2 + J_{l,m}^2 ) \, Y_{l,m} + J_{l,m} \, J_{l-1,m} \, Y_{l-2,m}
 \end{align}
with the constants given by
 \begin{align}
  J_{l,m} & = \sqrt{\frac{l^2-m^2}{4\,l^2-1}}
 \end{align}
\end{subequations}
Let us write down explicitly the equations for the three first coefficients in $\mathbf{B}$ and $\mathbf{D}$. The system of partial differential equations reads
\begin{subequations}
 \begin{align}
   \partial_r(\alpha^2\,\partial_r(r\,f_{1,0}^{B})) - \frac{2}{r} \, f_{1,0}^{B} & = \frac{6}{\sqrt{5}} \, \frac{\omega}{\varepsilon_0 \, c^2} \, f_{2,0}^{D} \\
 \label{eq:f30B}
 \partial_r(\alpha^2\,\partial_r(r\,f_{3,0}^{B})) - \frac{12}{r} \, f_{3,0}^{B} & = \frac{1}{\sqrt{7}} \, \frac{\omega}{\varepsilon_0 \, c^2} \, \left[ - 18 \, \sqrt{\frac{2}{5}} \, f_{2,0}^{D} + 4 \, \sqrt{15} \, f_{4,0}^{D} \right] \\
 \partial_r(\alpha^2\,\partial_r(r\,f_{5,0}^{B})) - \frac{30}{r} \, f_{5,0}^{B} & = \frac{2}{\sqrt{11}} \, \frac{\omega}{\varepsilon_0 \, c^2} \, \left[ - 5 \, \sqrt{6} \, f_{4,0}^{D} + 9 \, \sqrt{\frac{35}{13}} \, f_{6,0}^{D} \right] \\
\label{eq:f20D}
 \partial_r(\alpha^2\,\partial_r(r\,f_{2,0}^{D})) - \frac{6}{r} \, f_{2,0}^{D} & = \frac{6}{\sqrt{5}} \, \varepsilon_0 \, \omega \, \left[ f_{1,0}^{B} - 3 \sqrt{\frac{2}{7}} \, f_{3,0}^{B} \right] \\
\label{eq:f40D}
 \partial_r(\alpha^2\,\partial_r(r\,f_{4,0}^{D})) - \frac{20}{r} \, f_{4,0}^{D} & = 2 \, \sqrt{3} \, \varepsilon_0 \, \omega \, \left[ 2 \, \sqrt{\frac{5}{7}} \, f_{3,0}^{B} - 5 \, \sqrt{\frac{2}{11}} \, f_{5,0}^{B} \right]  \\
 \partial_r(\alpha^2\,\partial_r(r\,f_{6,0}^{D})) - \frac{42}{r} \, f_{6,0}^{D} & = 18 \, \sqrt{\frac{35}{143}} \, \varepsilon_0 \, \omega \, f_{5,0}^{B}
 \end{align}
\end{subequations}
The associated boundary conditions are
\begin{subequations}
\label{eq:CLfD}
\begin{align}
 \sqrt{\frac{2}{5}} \, \alpha^2 \, \partial_r(r\,f_{2,0}^{D}) & = 
 \varepsilon_0 \, r \, \tilde{\omega} \, \left[ \frac{2\,\sqrt{2}}{5} \, f_{1,0}^{B} - \frac{12}{5 \, \sqrt{7}} \, f_{3,0}^{B} \right] \\
 \alpha^2 \, \left[ \frac{2}{3}\,\sqrt{\frac{5}{7}} \, \partial_r(r\,f_{4,0}^{D}) - \sqrt{\frac{6}{35}} \, \partial_r(r\,f_{2,0}^{D}) \right] & = 
 \varepsilon_0 \, r \, \tilde{\omega} \, \left[ \frac{44}{15\,\sqrt{3}} \, f_{3,0}^{B} - \frac{2}{5} \, \sqrt{\frac{6}{7}} \, f_{1,0}^{B} - \frac{20}{3}\,\sqrt{\frac{10}{231}} \, f_{5,0}^{B} \right] \\
 \alpha^2 \, \left[ \sqrt{\frac{42}{143}} \, \partial_r(r\,f_{6,0}^{D}) - \frac{2}{3}\,\sqrt{\frac{5}{11}} \, \partial_r(r\,f_{4,0}^{D}) \right] & = 
 \varepsilon_0 \, r \, \tilde{\omega} \, \left[ \frac{58}{39} \, \sqrt{\frac{10}{3}} \, f_{5,0}^{B} - \frac{40}{3\,\sqrt{231}} \, f_{3,0}^{B} \right]
\end{align}
\end{subequations}
again where quantities have to be evaluated on the neutron star surface, at $r=R$.
We emphasize that the magnetic field at the neutron star surface is exactly matched to the expression for the general-relativistic static dipole, eq.~(\ref{eq:DipoleSchwarzf10}). All other multipole fields $f_{l,0}^{B}$ with $l\neq1$ vanish at $r=R$ by our definition.

\subsubsection{Numerical integration}

The computation of the electromagnetic field in vacuum has been reduced to a system of linear ordinary differential equations of second order. Moreover, it is a boundary value problem to be solved in a semi-infinite interval, from $r=R$ to $r=+\infty$. Several different techniques exist to treat such a system. We choose to employ spectral methods, expanding the unknown functions onto special basis functions. According to \cite{Boyd2001}, dealing with rational Chebyshev functions~$TL_k(y)$ is a judicious choice for the interval~$[R,+\infty[$. These functions are defined by
\begin{subequations}
 \begin{align}
  TL_k(y) & = T_k (x) \\
  x & = \frac{y-L}{y+L} \\
  y & = r - R
 \end{align}
\end{subequations}
where $T_k$ are the Chebyshev polynomials of order~$k$ and $y\in[0,+\infty[$. $L$ is a scaling parameter which should reproduce the characteristic length of the problem. We choose $L=R$. Any radial function~$f(r)$ is therefore expanded into a finite number of $N_r$~terms such that
\begin{equation}
 f(r) = \sum_{k=0}^{N_r-1} f_k \, TL_k(y(r))
\end{equation}
The inner and outer boundary conditions for the magnetic field coefficients~$f^B_{l,m}$ are expressed as
\begin{subequations}
 \begin{align}
  \label{eq:CLf1}
  \sum_{k=0}^{N_r-1} f_k & = 0 \\
  \sum_{k=0}^{N_r-1} (-1)^k \, f_k & = f(R) 
 \end{align}
\end{subequations}
Actually, all the $f(R)$ vanish except for the dipole $f^B_{1,0}(R)$, recall that we strictly impose a dipolar magnetic field on the neutron star surface. The outer boundary conditions for the electric field coefficients~$f^D_{l,m}$ are the same as eq.~(\ref{eq:CLf1}). The inner boundary conditions are different because we enforce conditions on the derivative, see eq.~(\ref{eq:CLfD}), but it remains a relation involving linear terms in the expansion coefficients. We use what \cite{Boyd2001} calls the boundary-bordering method leading to a linear algebra system to be solved. For more details on spectral and pseudo-spectral methods, see for instance also \cite{2006spme.book.....C}. Technically, we use Mathematica~9 to compute the matrix coefficients and invert the system to find the expansion coefficients.

\begin{figure}
  \centering
  \includegraphics[width=0.5\textwidth]{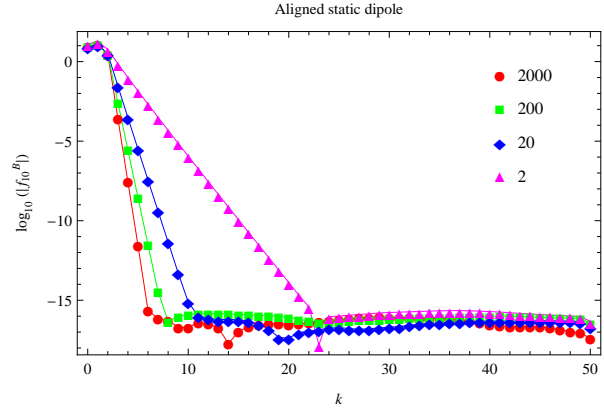}
  \caption{Absolute value of the coefficients of the rational Chebyshev expansion of the magnetic field functions $f_{1,0}^{B}$ for the static aligned dipole. $k$ corresponds to the order of the k-th rational Chebyshev function and the numbers in the legend depict the ratio~$R/\Rs$.}
\label{fig:fb10_align}
\end{figure}

\begin{figure}
  \centering
  \includegraphics[width=0.5\textwidth]{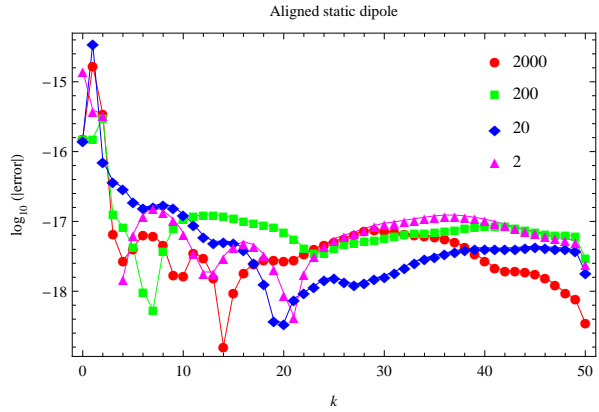}
  \caption{Absolute error of the numerical solution compared to the analytical expression for the magnetic field~$f_{1,0}^{B}$ for the static aligned dipole. $k$ corresponds to the order of the k-th rational Chebyshev function and the numbers in the legend depict the ratio~$R/\Rs$.}
\label{fig:fb10_align_erreur}
\end{figure}

For the subsequent numerical applications, we normalize the magnetic moment of the neutron star to unity, $\mu=1$. In order to demonstrate the accuracy of our spectral algorithm to solve the system of linear ordinary differential equations, we begin with the static aligned dipole. The neutron star radius is set to $R=\rlight/10$ although it is irrelevant for the static dipole case because there is no rotation and no scaling with~$\rlight$. The Schwarzschild radius is chosen with increasing value compared to the stellar radius, we take $R=\{2,20,200,2000\} \, \Rs$. The absolute value of the expansion coefficients of the function~$f_{1,0}^{B}$ are shown in fig.~\ref{fig:fb10_align} with a number of collocations points $N_{\rm r}=51$. For any value of the compactness parameter~$\Xi$, we get the prescribed 15~digits of accuracy, although that for compactness close to unity, we need more coefficients. Indeed, for very low compactness $\Xi=1/2000$, red curve with full circles, less than ten coefficients are required to get full accuracy. The same remark holds for any $\Xi\ll1$. For the typical compactness of a neutron star, $\Xi=0.5$, magenta curve with full triangles, we need almost 25~coefficients to achieve the required accuracy. Nevertheless, the spectral convergence of our computation is clearly identified by the exponential decrease of the magnitude of the highest-order coefficients within an accuracy of 15~digits. The other question to address is the efficiency of our spectral algorithm to reproduce the exact solution depending on the strength of deviation from the flat space-time metric. In order to check the correctness of these coefficients, we have to compare them with the expansion coefficients of the analytical solution given by eq.~(\ref{eq:DipoleSchwarzf10}).
To do this, we define the absolute error between the analytical~$f_{1,0}^{B({\rm dip})}$ and the numerical~$f_{1,0}^{B({\rm num})}$ solution by
\begin{equation}
 \label{eq:ErreurAbs}
 \textrm{error} (f_{1,0}^{B}) = \left| \frac{f_{1,0}^{B({\rm dip})} - f_{1,0}^{B{(\rm num})}}{{\rm max}(f_{1,0}^{B({\rm dip})})} \right|
\end{equation}
This error is plotted in fig.~\ref{fig:fb10_align_erreur} and shows a perfect match between both solutions, within the numerical accuracy. We reach 15~digits of significance for the relevant coefficients, those which are not zero numerically. This explains the decreasing number of significant digits when the coefficients are close to zero. They are meaningless.

This first example demonstrates the very high accuracy obtainable by our spectral method. Next, we pursue with the rotating aligned dipole. Rotation combined with frame dragging effects will produce higher order multipole coefficients which to first order depend linearly on the spin parameter~$a$. We already gave an approximate analytical solution to the lowest order, i.e. the induced electric field without taking into account the perturbation in the magnetic field. Nevertheless with our numerical integration procedure, we are able to give solutions to any order in the multipole moments~$l$. We therefore proceed in an increasing order of complexity. Starting with only the two functions $f_{1,0}^{B}$ and $f_{2,0}^{D}$ to the lowest approximation, corresponding to the magnetic dipole and to the electric quadrupole, we then successively add the couple $(f_{3,0}^{B},f_{4,0}^{D})$ and conclude with two more functions $(f_{5,0}^{B},f_{6,0}^{D})$. Consequently, we can quantitatively estimate the contribution to the electromagnetic field from higher multipoles other than dipole and quadrupole.

\begin{figure}
  \centering
 \includegraphics[width=0.5\textwidth]{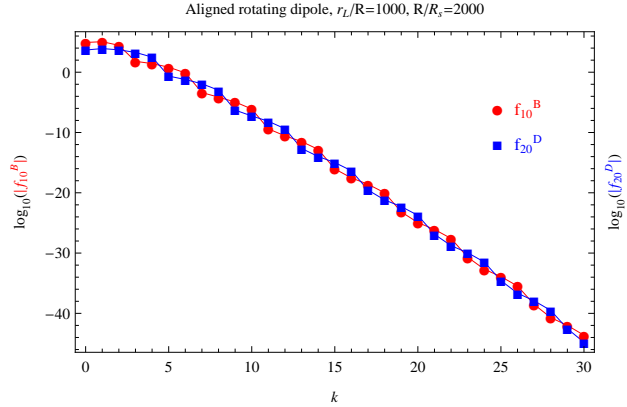}
  \caption{Absolute value of the coefficients of the rational Chebyshev expansion of the magnetic and electric field functions, $f_{1,0}^{B}$ in red circles and $f_{2,0}^{D}$ in blue squares, for the aligned rotating dipole. The parameters are $R=2000\,\Rs$ and $\rlight=1000\,R$.}
\label{fig:f_align_rot_1_r1000_rs2000}
\end{figure}

\begin{figure}
  \centering
 \includegraphics[width=0.5\textwidth]{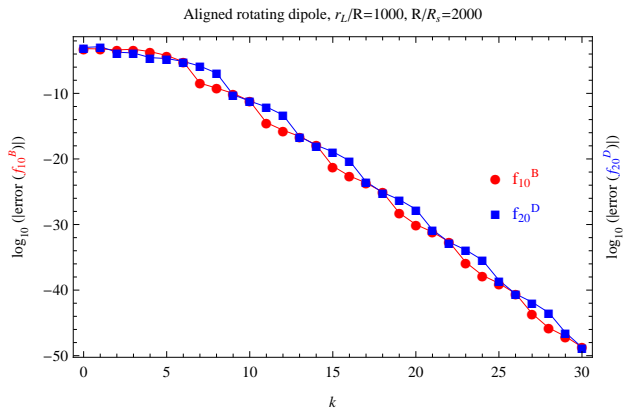}
  \caption{Difference of the numerical solution compared to the first order analytical approximate expression for the magnetic field~$f_{1,0}^{B({\rm dip})}$ in red circles and for the electric field~$f_{2,0}^{D({\rm quad})}$ in blue squares. The parameters are $R=2000\,\Rs$ and $\rlight=1000\,R$.}
\label{fig:erreur_align_rot_1_r1000_rs2000}
\end{figure}

\begin{figure}
  \centering
 \includegraphics[width=0.5\textwidth]{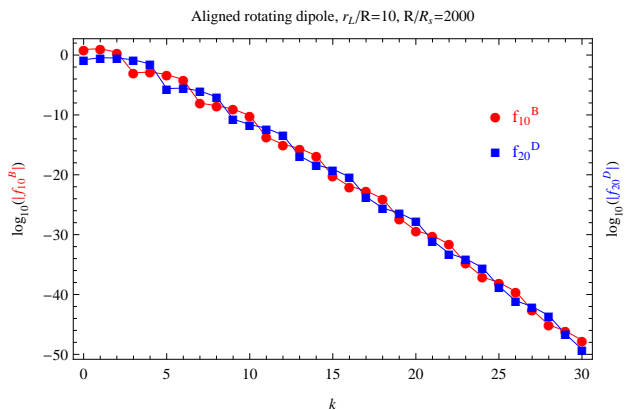}
  \caption{Absolute value of the coefficients of the rational Chebyshev expansion of the magnetic field and electric field functions, $f_{1,0}^{B}$ in red circles and $f_{2,0}^{D}$ in blue squares, for the aligned rotating dipole. The parameters are $R=2000\,\Rs$ and $\rlight=10\,R$.}
\label{fig:f_align_rot_1_r10_rs2000}
\end{figure}

\begin{figure}
  \centering
 \includegraphics[width=0.5\textwidth]{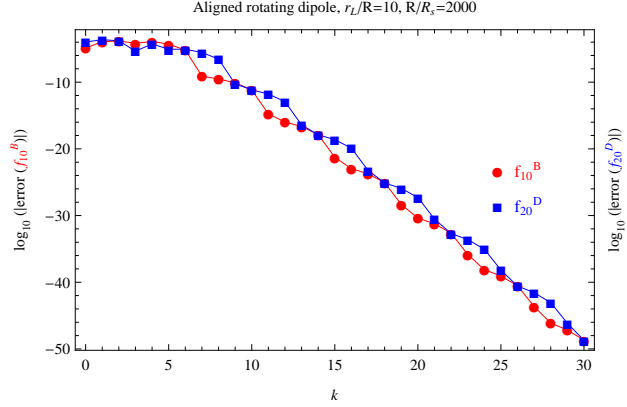}
  \caption{Difference of the numerical solution compared to the first order analytical approximate expression for the magnetic field~$f_{1,0}^{B({\rm dip})}$ in red circles and for the electric field~$f_{2,0}^{D({\rm quad})}$ in blue squares. The parameters are $R=2000\,\Rs$ and $\rlight=10\,R$.}
\label{fig:erreur_align_rot_1_r10_rs2000}
\end{figure}

\begin{figure}
  \centering
 \includegraphics[width=0.5\textwidth]{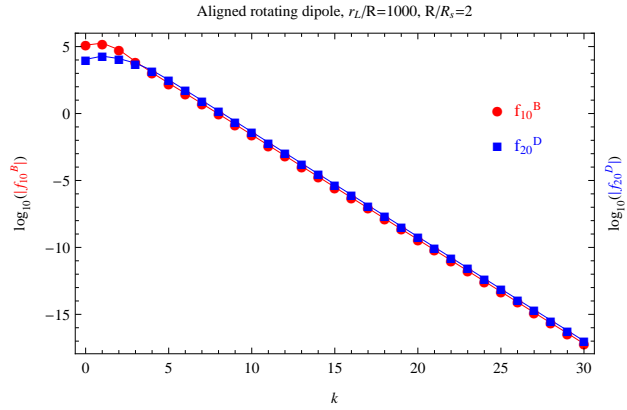}
  \caption{Absolute value of the coefficients of the rational Chebyshev expansion of the magnetic field and electric field functions, $f_{1,0}^{B}$ in red circles and $f_{2,0}^{D}$ in blue squares, for the aligned rotating dipole. The parameters are $R=2\,\Rs$ and $\rlight=1000\,R$.}
\label{fig:f_align_rot_1_r1000_rs2}
\end{figure}

\begin{figure}
  \centering
 \includegraphics[width=0.5\textwidth]{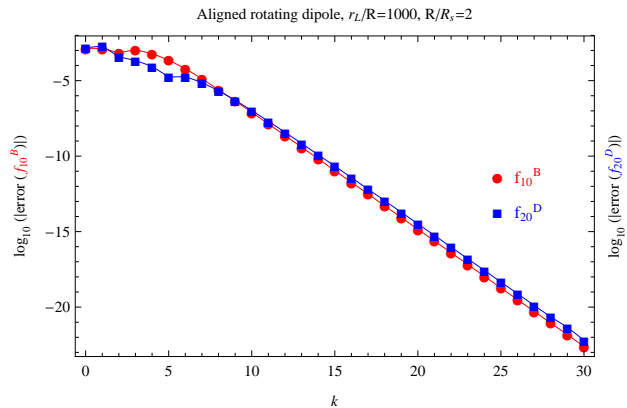}
  \caption{Difference of the numerical solution compared to the first order analytical approximate expression for the magnetic field~$f_{1,0}^{B({\rm dip})}$ in red circles and for the electric field~$f_{2,0}^{D({\rm quad})}$ in blue squares. The parameters are $R=2\,\Rs$ and $\rlight=1000\,R$.}
\label{fig:erreur_align_rot_1_r1000_rs2}
\end{figure}

\begin{figure}
  \centering
 \includegraphics[width=0.5\textwidth]{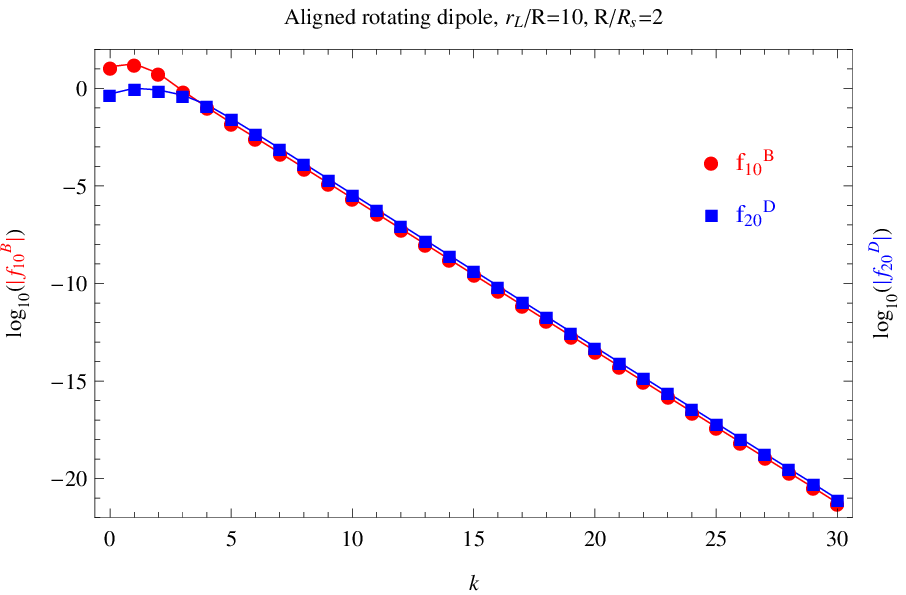}
  \caption{Absolute value of the coefficients of the rational Chebyshev expansion of the magnetic field and electric field functions, $f_{1,0}^{B}$ in red circles and $f_{2,0}^{D}$ in blue squares, for the aligned rotating dipole. The parameters are $R=2\,\Rs$ and $\rlight=10\,R$.}
\label{fig:f_align_rot_1_r10_rs2}
\end{figure}

\begin{figure}
  \centering
 \includegraphics[width=0.5\textwidth]{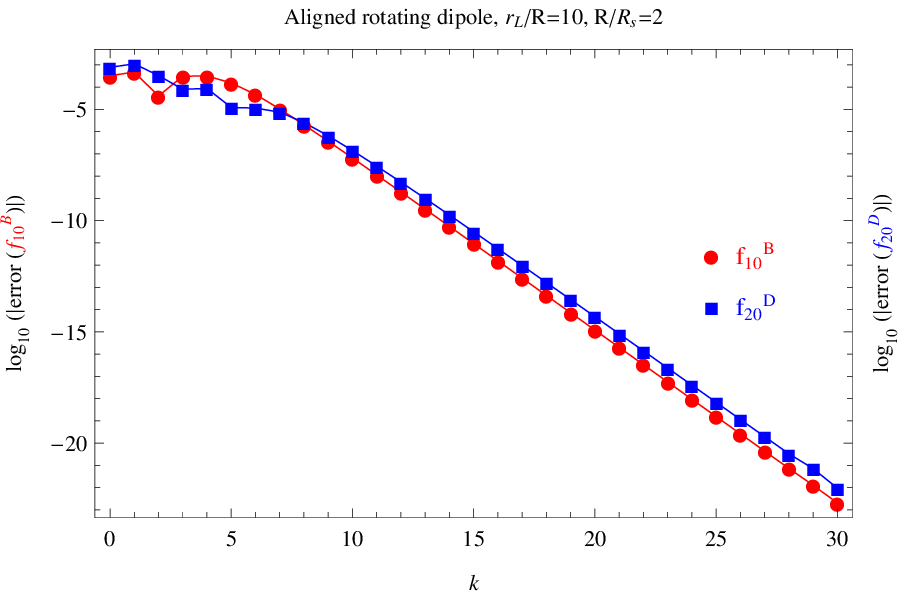}
  \caption{Difference of the numerical solution compared to the first order analytical approximate expression for the magnetic field~$f_{1,0}^{B({\rm dip})}$ in red circles and for the electric field~$f_{2,0}^{D({\rm quad})}$ in blue squares. The parameters are $R=2\,\Rs$ and $\rlight=10\,R$.}
\label{fig:erreur_align_rot_1_r10_rs2}
\end{figure}

\begin{figure}
  \centering
 \includegraphics[width=0.5\textwidth]{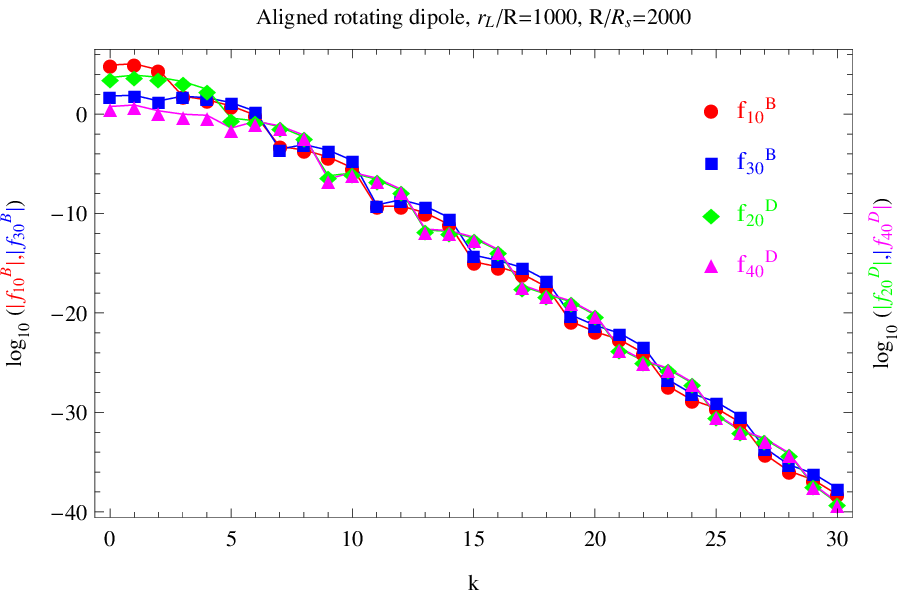}
  \caption{Absolute value of the coefficients of the rational Chebyshev expansion of the magnetic field and electric field functions $f_{1,0}^{B}, f_{3,0}^{B}$ and $f_{2,0}^{D}, f_{4,0}^{D}$ for the aligned rotating dipole. The parameters are $R=2000\,\Rs$ and $\rlight=1000\,R$.}
\label{fig:f_align_rot_2_r1000_rs2000}
\end{figure}

\begin{figure}
  \centering
 \includegraphics[width=0.5\textwidth]{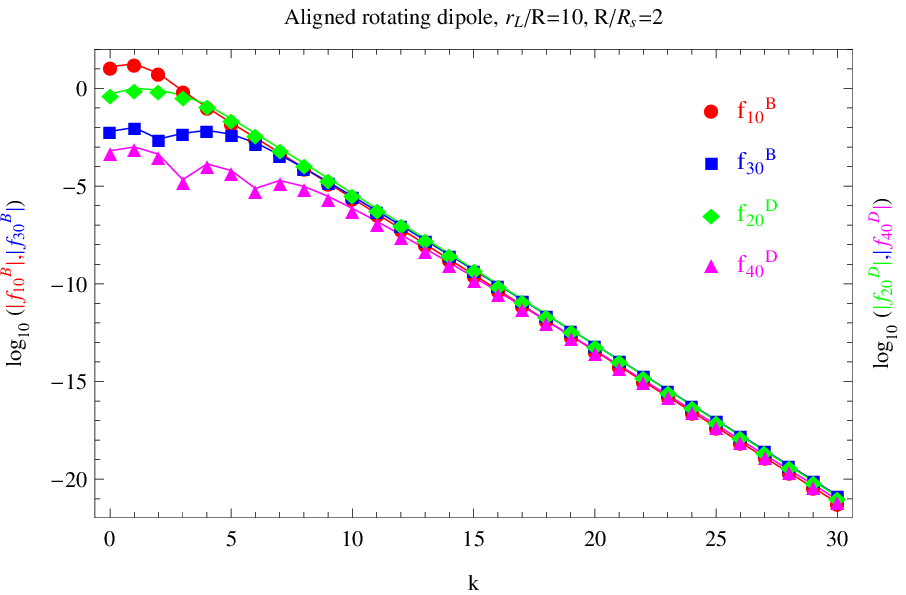}
  \caption{Absolute value of the coefficients of the rational Chebyshev expansion of the magnetic field and electric field functions $f_{1,0}^{B}, f_{3,0}^{B}$ and $f_{2,0}^{D}, f_{4,0}^{D}$ for the aligned rotating dipole. The parameters are $R=2\,\Rs$ and $\rlight=10\,R$.}
\label{fig:f_align_rot_2_r10_rs2}
\end{figure}

\begin{figure}
  \centering
 \includegraphics[width=0.5\textwidth]{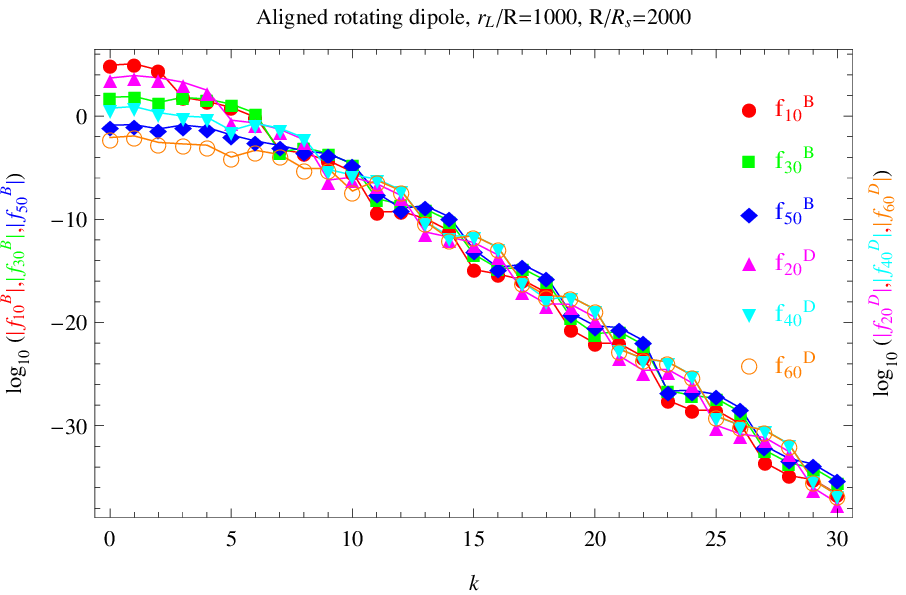}
  \caption{Absolute value of the coefficients of the rational Chebyshev expansion of the magnetic field and electric field functions $f_{1,0}^{B}, f_{3,0}^{B}, f_{5,0}^{B}$ and $f_{2,0}^{D}, f_{4,0}^{D}, f_{6,0}^{D}$ for the aligned rotating dipole.}
\label{fig:f_align_rot_3_r1000_rs2000}
\end{figure}

\begin{figure}
  \centering
 \includegraphics[width=0.5\textwidth]{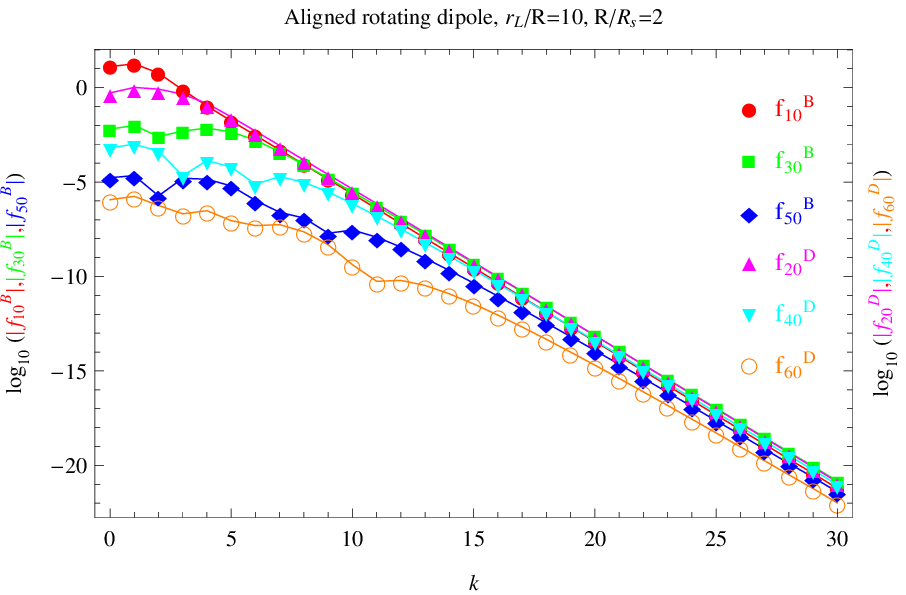}
  \caption{Absolute value of the coefficients of the rational Chebyshev expansion of the magnetic field and electric field functions $f_{1,0}^{B}, f_{3,0}^{B}, f_{5,0}^{B}$ and $f_{2,0}^{D}, f_{4,0}^{D}, f_{6,0}^{D}$ for the aligned rotating dipole.}
\label{fig:f_align_rot_3_r10_rs2}
\end{figure}

\begin{figure}
  \centering
\begin{tabular}{cc}
 \includegraphics[width=0.5\textwidth]{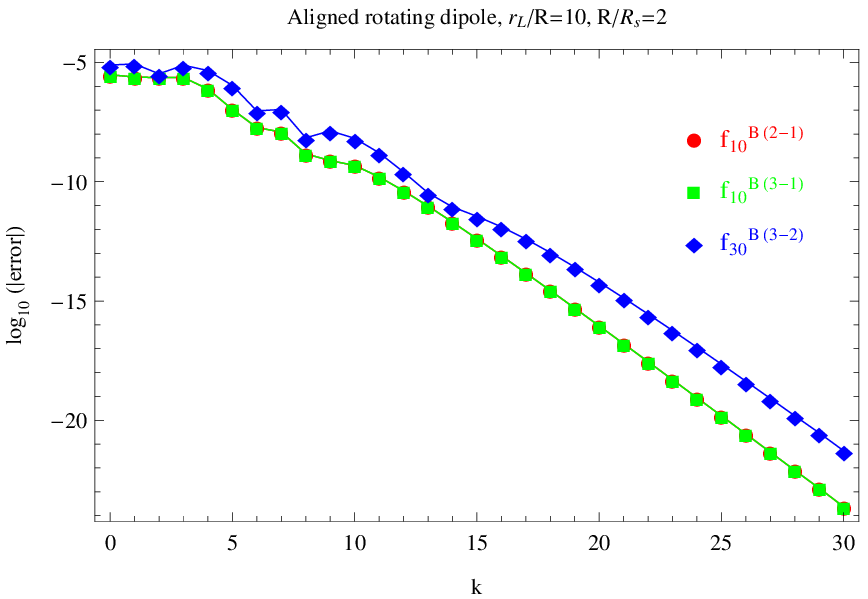} &
 \includegraphics[width=0.5\textwidth]{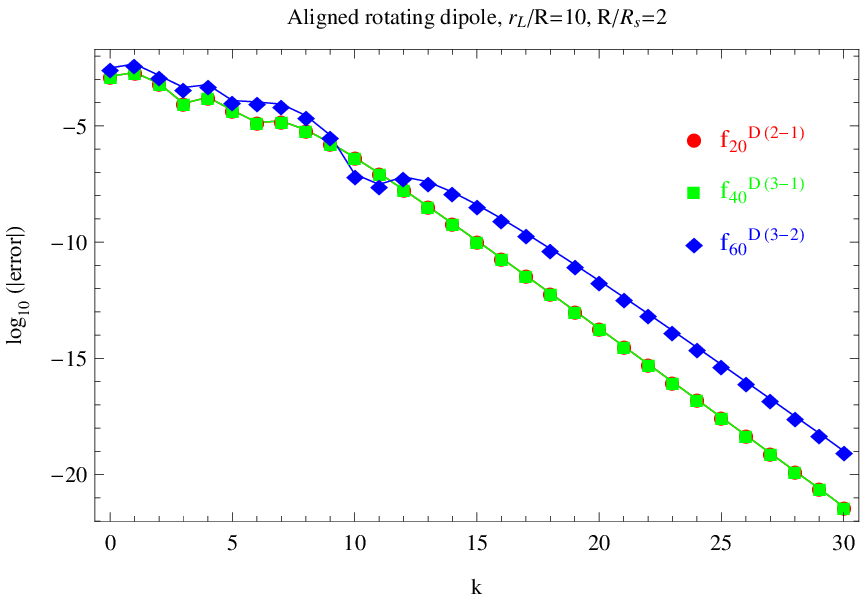} 
\end{tabular}
  \caption{Discrepancy between the coefficients of the rational Chebyshev expansion of the magnetic and electric field functions $f\{_{1,0}^{B}, f_{3,0}^{B}\}$ and $\{f_{2,0}^{D}, f_{4,0}^{D}\}$ for the aligned rotating dipole depending on the number of multipoles used in the expansion. $(2-1)$ means comparison between an expansion with four multipoles, two for $B$ and two for $D$, and two multipoles, magnetic dipole and electric quadrupole. $(3-1)$ means comparison between an expansion with six multipoles, three for $B$ and three for $D$, and two multipoles, magnetic dipole and electric quadrupole. $(3-2)$ means comparison between an expansion with six multipoles, three for $B$ and three for $D$, and four multipoles. The case with only a magnetic dipole and an electric quadrupole is excluded because it contains only $f_{1,0}^{B}$ and $f_{2,0}^{D}$.}
\label{fig:erreur_multipole_r10_rs2}
\end{figure}

\begin{figure}
  \centering
\begin{tabular}{cc}
 \includegraphics[width=0.5\textwidth]{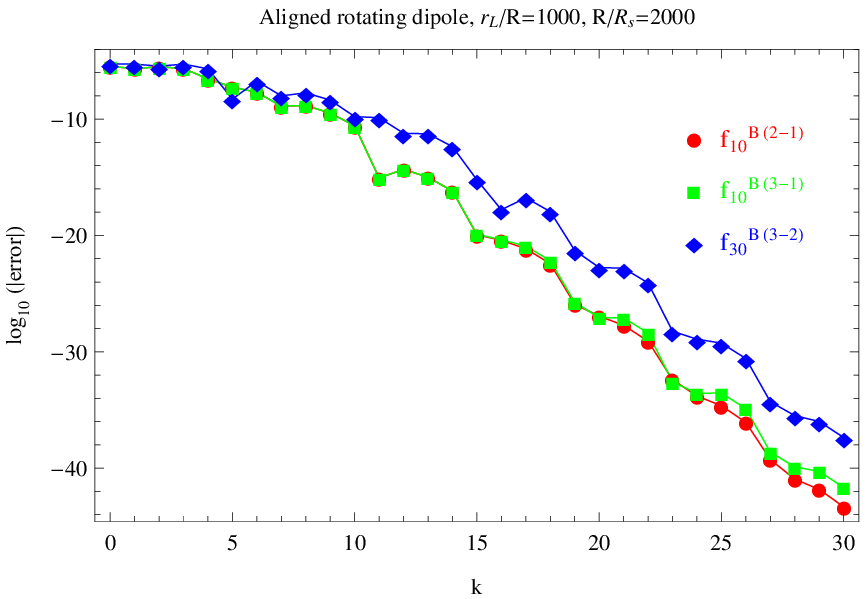} &
 \includegraphics[width=0.5\textwidth]{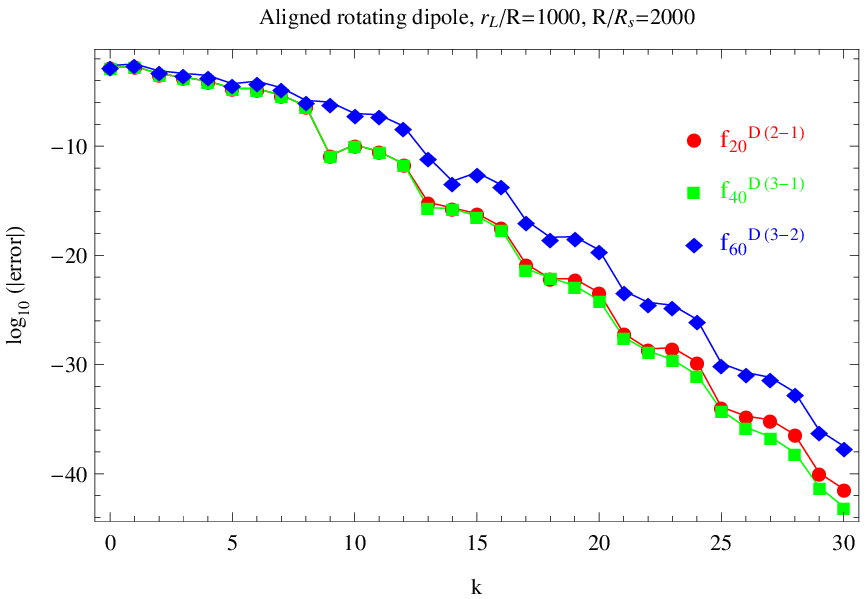} 
\end{tabular}
  \caption{Same as fig.~\ref{fig:erreur_multipole_r10_rs2} but for a weakly rotating and non compact star.}
\label{fig:erreur_multipole_r1000_rs2000}
\end{figure}

We performed different sets of calculation by combining slow and fast rotation $\rlight=\{10,1000\}\,R$ with low and high compactness $R=\{2,2000\}\,\Rs$ with normalized magnetic moment $\mu=1$. We start with a very slowly rotating dipole for which $\rlight=1000\,R$ and a low compactness $R=2000\,\Rs$ in order to look for small perturbations of the electric field induced by frame dragging effects. We can therefore compare the approximate analytical expressions with the more accurate numerical one. The absolute value of the rational Chebyshev coefficients of the lowest order approximation are shown in fig.~\ref{fig:f_align_rot_1_r1000_rs2000} for $f_{1,0}^{B}$ and $f_{2,0}^{D}$. Spectral convergence is achieved as expected. The discrepancy between the analytical solution and the numerical computation are small, less than $10^{-3}$, the absolute error between both sets of coefficients is close to zero as can be seen in fig.~\ref{fig:erreur_align_rot_1_r1000_rs2000}.
In a second set of calculations, we increased the frame-dragging effects by taking $\rlight=10\,R$ and $R=2000\,\Rs$. The absolute value of the rational Chebyshev coefficients of the lowest order approximation are shown in fig.~\ref{fig:f_align_rot_1_r10_rs2000} for $f_{1,0}^{B}$ and $f_{2,0}^{D}$. Spectral convergence is achieved as expected. Here also the absolute discrepancy between both sets of coefficients is close to zero as can be seen in fig.~\ref{fig:erreur_align_rot_1_r10_rs2000}.
In a third set of calculations, we increased the compactness by taking $\rlight=1000\,R$ and $R=2\,\Rs$. These values are typical for radio pulsars. The absolute value of the rational Chebyshev coefficients of the lowest order approximation are shown in fig.~\ref{fig:f_align_rot_1_r1000_rs2} for $f_{1,0}^{B}$ and $f_{2,0}^{D}$. Spectral convergence is achieved as expected. Here also the absolute discrepancy between both sets of coefficients is close to zero as can be seen in fig.~\ref{fig:erreur_align_rot_1_r1000_rs2}.
In a last set of calculations, we increased the rotation frequency by taking $\rlight=10\,R$ and $R=2\,\Rs$. These values are typical for millisecond pulsars. The absolute value of the rational Chebyshev coefficients of the lowest order approximation are shown in fig.~\ref{fig:f_align_rot_1_r10_rs2} for $f_{1,0}^{B}$ and $f_{2,0}^{D}$. Spectral convergence is achieved as expected. The absolute discrepancy is shown in fig.~\ref{fig:erreur_align_rot_1_r10_rs2}
Next, we go on in this section about the aligned rotator by computing higher order multipoles $l=\{3,4\}$ to demonstrate that they are several orders of magnetic less than the magnetic dipolar and electric quadrupolar moment. Results are shown in fig.~\ref{fig:f_align_rot_2_r1000_rs2000} for two more multipoles with a slowly rotating non compact star. The same for a rapidly rotating neutron star is shown in fig.~\ref{fig:f_align_rot_2_r10_rs2}.
We conclude this section by computing even higher order multipoles $l=\{5,6\}$ to demonstrate that they are also several orders of magnetic less than the lower multipolar moments. For a total of 6 multipoles, we get the coefficients represented in fig.~\ref{fig:f_align_rot_3_r1000_rs2000} for the slowly rotating non compact star and for a rapidly rotating neutron star in fig.~\ref{fig:f_align_rot_3_r10_rs2}.
The dipolar magnetic field as well as the electric quadrupolar field are not significantly affected by the higher multipolar fields. Indeed, we show the discrepancy in the expansion coefficients in fig.~\ref{fig:erreur_multipole_r10_rs2}. We first compare the dipole magnetic field quadrupole electric field expansion versus a dipole plus octupole~$l=3$ expansion of the magnetic field and a quadrupole plus $l=4$ electric fields, denoted by $f_{1,0}^{B(2-1)}$ and $f_{2,0}^{D(2-1)}$. The same can be performed with a threefold expansion for both fields and denoted by $f_{1,0}^{B(3-1)}$ and $f_{2,0}^{D(3-1)}$. Higher order multipoles can also be compared by inspection of~$f_{3,0}^{B(3-2)}$ and $f_{4,0}^{D(3-2)}$. Comparison with the lowest order expansion is not possible because this approximate solution does not contain neither $f_{3,0}^{B}$ nor $f_{4,0}^{D}$. We conclude from the plots in fig.~\ref{fig:erreur_multipole_r10_rs2} that the discrepancy in the expansion coefficients is not relevant. In other words, adding higher multipoles will not significantly perturb the lower expansion coefficients. For an almost non rotating and non compact star, the discrepancies are shown in fig.~\ref{fig:erreur_multipole_r1000_rs2000}. They are weaker than in the previous case.

\section{ELECTROMAGNETIC FIELD OF AN ORTHOGONAL DIPOLE}
\label{sec:Orthogonal}

In this last section, we investigate the orthogonal rotator in vacuum as a generalization of the Deutsch solution. We first check that we retrieve the static perpendicular dipole magnetic field in a Schwarzschild space-time. We resume the section with the perpendicular rotating dipole to modest numerical accuracy. 

\subsection{Static dipole}

The orthogonal static dipole follows the same lines as those for the aligned static dipole. Far from the neutron star, we expect to retrieve the flat space-time expression so we develop the magnetic field according to
\begin{equation}
  \label{eq:BstaticOrtho1}
  \mathbf B = \rot ( f_{1,1}^B \, \mathbf \Phi_{1,1} )
\end{equation}
with the boundary condition at infinity such that the magnetic field becomes
\begin{equation}
  \label{eq:BstaticOrthoPlat}
  \mathbf B = Re \left[ \sqrt{\frac{16\,\pi}{3}} \, \frac{\mu_0\,\mu}{4\,\pi} \rot ( \frac{\mathbf \Phi_{1,1}}{r^2} ) \right] 
\end{equation}
This corresponds to the boundary conditions at infinity
\begin{equation}
  \label{eq:BstaticOrtho2}
  \lim\limits_{r\to+\infty} f_{1,1}^B = \sqrt{\frac{16\,\pi}{3}} \, \frac{\mu_0\,\mu}{4\,\pi\,r^2}
\end{equation}
We take as a definition for an orthogonal dipole the presence of only one spherical harmonic, namely $(l,m)=(1,1)$. The procedure then follows exactly the same lines as for the static aligned dipole. We refer to this case for more details about the calculations.
It is then straightforward to show that the scalar function~$f_{1,1}^B$ must satisfy the same equation as $f_{1,0}^B$, namely
\begin{equation}
  \label{eq:Laplacef11}
  \partial_r(\alpha^2\,\partial_r(r\,f_{1,1}^B)) - \frac{2}{r} \, f_{1,1}^B = 0
\end{equation}
The exact solution for the dipole magnetic field which asymptotes to the flat dipole is given by
\begin{equation}
  \label{eq:DipoleSchwarzf11}
  f_{1,1}^{B({\rm dip})} = - \sqrt{\frac{16\,\pi}{3}} \, \frac{3\,\mu_0\,\mu\,r}{4\,\pi\,R_s^3} \, \left[ {\rm ln}\,\alpha^2 + \frac{R_s}{r} + \frac{R_s^2}{2\,r^2}\right]
\end{equation}
which is related to the aligned solution by
\begin{equation}
 f_{1,1}^{B({\rm dip})} = - \sqrt{2} \, f_{1,0}^{B({\rm dip})}
\end{equation}
The magnetic field components are
\begin{subequations}
\begin{align}
  \label{eq:MagneticStaticOrtho}
  B^{\hat r} = & - 6 \, \left[ {\rm ln} \, \alpha^2 + \frac{R_s}{r} + \frac{R_s^2}{2\,r^2} \right] \, \frac{\mu_0\,\mu\,\sin\vartheta\,\cos\varphi}{4\,\pi\,R_s^3} \\
  B^{\hat \vartheta} = & - 3 \, \left[ 2 \, \alpha \, {\rm ln} \,
    \alpha^2 + \frac{R_s}{r} \, \frac{2\,r-R_s}{\sqrt{r\,(r-R_s)}}
  \right] \, \frac{\mu_0\,\mu\,\cos\vartheta\,\cos\varphi}{4\,\pi\,R_s^3} \\
  B^{\hat \varphi} = & + 3 \, \left[ 2 \, \alpha \, {\rm ln} \,
    \alpha^2 + \frac{R_s}{r} \, \frac{2\,r-R_s}{\sqrt{r\,(r-R_s)}}
  \right] \, \frac{\mu_0\,\mu\,\sin\varphi}{4\,\pi\,R_s^3} 
\end{align}
\end{subequations}
Corrections to first order compared to flat space-time are
\begin{subequations}
\begin{align}
  \label{eq:CorrectionOrdre1}
    B^{\hat r} = & \frac{2\,\mu_0\,\mu\,\sin\vartheta\,\cos\varphi}{4\,\pi\,r^3} \, \left[ 1 + \frac{3}{4} \, \frac{R_s}{r} + o\left( \frac{R_s}{r} \right) \right] \\
  B^{\hat \vartheta} = & - \frac{\mu_0\,\mu\,\cos\vartheta\,\cos\varphi}{4\,\pi\,r^3} \, \left[ 1 + \frac{R_s}{r} + o\left( \frac{R_s}{r} \right) \right] \\
  B^{\hat \varphi} = & \frac{\mu_0\,\mu\,\sin\varphi}{4\,\pi\,r^3} \, \left[ 1 + \frac{R_s}{r} + o\left( \frac{R_s}{r} \right) \right]
\end{align} 
\end{subequations}
We now switch to the most interesting case, the general relativistic orthogonal rotating dipole in vacuum.

\subsection{Stationary rotator}

\subsubsection{General formalism to any order}

We next look for the stationary solution to Maxwell equations in curved vacuum space. In this vacuum, the fields $\mathbf{D}$ and $\mathbf{B}$ are divergencelessness. We therefore expand them according to the most general prescription
\begin{subequations}
\begin{align}
  \label{eq:Decomposition_HSV_div_0_D}
  \mathbf{D}(r,\vartheta,\varphi,t) = & \sum_{l=1}^\infty\sum_{m=-l}^l \left( \rot [f^D_{l,m}(r,t) \, \mathbf{\Phi}_{l,m}] + g^D_{l,m}(r,t) \, \mathbf{\Phi}_{l,m} \right) \\
  \label{eq:Decomposition_HSV_div_0_B}
  \mathbf{B}(r,\vartheta,\varphi,t) = & \sum_{l=1}^\infty\sum_{m=-l}^l \left( \rot [f^B_{l,m}(r,t) \, \mathbf{\Phi}_{l,m}] + g^B_{l,m}(r,t) \, \mathbf{\Phi}_{l,m} \right)
\end{align} 
\end{subequations}
We extended the method outlined by \cite{2012MNRAS.424..605P} and employ complex quantities. Therefore, the time-dependent part is proportional to $e^{-i\,m\,\Omega\,t}$. Remember that the divergence-free property of the electromagnetic field is insured by construction, it is a consequence of the above expansion, eqs.~(\ref{eq:Decomposition_HSV_div_0_D})-(\ref{eq:Decomposition_HSV_div_0_B}). The remaining Maxwell equations involving the curl are satisfied if and only if the $f^{D}_{l,m}$ are solutions to the second order linear partial differential equation
\begin{subequations}
\begin{multline}
 \alpha \, \mathcal{R}_l[f^{D}_{l,m}] = - i \, \varepsilon_0 \, m \, ( \Omega - \omega ) \, g^{B}_{l,m} + \\
 3 \, \varepsilon_0 \, \alpha \, \frac{\omega}{r} \left[ f^{B}_{l-1,m} \, \sqrt{(l-1)(l+1)} \, J_{l,m} - f^{B}_{l+1,m} \, \sqrt{l\,(l+2)} \, J_{l+1,m} \right] 
\end{multline}
and similarly for the coefficients~$f^{B}_{l,m}$
\begin{multline}
 \alpha \, \mathcal{R}_l[f^{B}_{l,m}] = i \, \mu_0 \, m \, ( \Omega - \omega ) \, g^{D}_{l,m} -\\
 3 \, \mu_0 \, \alpha \, \frac{\omega}{r} \left[ f^{D}_{l-1,m} \, \sqrt{(l-1)(l+1)} \, J_{l,m} - f^{D}_{l+1,m} \, \sqrt{l\,(l+2)} \, J_{l+1,m} \right] 
\end{multline}
To derive these expressions, we put the expansions eqs.~(\ref{eq:Decomposition_HSV_div_0_D})-(\ref{eq:Decomposition_HSV_div_0_B}) into eqs.~(\ref{eq:Maxwell2})-(\ref{eq:Maxwell4}), then project onto the $\mathbf{\Phi}_{lm}$ and used identities from the appendix~\ref{sec:framedragging}. Moreover, there exists a simple algebraic relation between $g^{D}_{l,m}$ and $f^{B}_{l,m}$ on one side, and between $g^{B}_{l,m}$ and $f^{D}_{l,m}$ on the other side. We find
\begin{align}
\label{eq:gDvsfB}
 \alpha \, g^{D}_{l,m} = & + i \, \varepsilon_0 \, m \, \tilde{\omega} \, f^{B}_{l,m} \\
 \alpha \, g^{B}_{l,m} = & - i \,         \mu_0 \, m \, \tilde{\omega} \, f^{D}_{l,m} 
\end{align}
\end{subequations}
obtained by projection of the same equations but now onto $\er$.
All these relations can be summarized in two inhomogeneous Helmholtz equations for the electric field~$f^{D}_{l,m}$
\begin{subequations}
\label{eq:Helmholtz}
\begin{multline}
\label{eq:HelmholtzD}
 \alpha^2 \, \mathcal{R}_l[f^{D}_{l,m}] + m^2 \, \frac{\tilde\omega^2}{c^2} \, f^{D}_{l,m} = \\
 3 \, \varepsilon_0 \, \alpha^2 \, \frac{\omega}{r} \, \left[ f^{B}_{l-1,m} \, \sqrt{(l-1)(l+1)} \, J_{l,m} - f^{B}_{l+1,m} \, \sqrt{l\,(l+2)} \, J_{l+1,m} \right]
\end{multline}
and similarly for the magnetic field~$f^{B}_{l,m}$
\begin{multline}
\label{eq:HelmholtzB}
 \alpha^2 \, \mathcal{R}_l[f^{B}_{l,m}] + m^2 \, \frac{\tilde\omega^2}{c^2} \, f^{B}_{l,m} = \\
- 3 \, \mu_0 \, \alpha^2 \, \frac{\omega}{r} \, \left[ f^{D}_{l-1,m} \, \sqrt{(l-1)(l+1)} \, J_{l,m} - f^{D}_{l+1,m} \, \sqrt{l\,(l+2)} \, J_{l+1,m} \right] 
\end{multline}
\end{subequations}
The boundary conditions on the neutron star surface are imposed in the following way. Introducing the expansions eq.~(\ref{eq:Decomposition_HSV_div_0_D}) and eq.~(\ref{eq:Decomposition_HSV_div_0_B}) into eq.~(\ref{eq:CLD}), then projecting along~$\etheta$ and~$\ephi$ using the formula eq.~(\ref{eq:BETAVHS}) in appendix~\ref{app:HSV} we get the relation between the coefficients of $\mathbf D$ and $\mathbf B$ as
\begin{subequations}
  \begin{align}
    \sum_{l,m} i \, \frac{g^D_{l,m}}{\sqrt{l\,(l+1)}} \, \sin\vartheta \, \partial_\vartheta Y_{l,m} & = - \frac{m \, \alpha}{r\, \sqrt{l\,(l+1)}} \, \partial_r ( r \, f^D_{l,m}) \, Y_{l,m} \\
    \sum_{l,m} - \frac{\alpha}{r} \, \partial_r ( r \, f^D_{l,m}) \, \frac{\sin\vartheta}{\sqrt{l\,(l+1)}} \, \partial_\vartheta Y_{l,m} - i \, \frac{m}{\sqrt{l\,(l+1)}} \, g^D_{l,m} \, Y_{l,m} & = \varepsilon_0 \, \frac{\Omega-\omega}{\alpha} \, \sin^2\vartheta \, \sqrt{l\,(l+1)} \, f^B_{l,m} \, Y_{l,m} 
  \end{align}
\end{subequations}
This can be rearranged by indexation with the same $Y_{l,m}$ such that
\begin{subequations}
  \begin{align}
\label{eq:BC1}
    \sqrt{\frac{l-1}{l}} \, J_{l,m} \, g^D_{l-1,m} - \sqrt{\frac{l+2}{l+1}} \, J_{l+1,m} \, g^D_{l+1,m} = \frac{i\,m\,\alpha}{r\,\sqrt{l\,(l+1)}} \, \partial_r ( r \, f^D_{l,m}) & \\
\label{eq:BC2}
  \alpha^2 \, \sqrt{\frac{l+2}{l+1}} \, J_{l+1,m} \, \partial_r ( r \, f^D_{l+1,m}) - \alpha^2 \, \sqrt{\frac{l-1}{l}} \, J_{l,m} \, \partial_r ( r \, f^D_{l-1,m}) - i \, \frac{m \, \alpha \, r}{\sqrt{l\,(l+1)}} \, g^D_{l,m} & = \\
    \varepsilon_0 \, r \, \tilde{\omega} \, \left[ \sqrt{l\,(l+1)} \, ( 1 - J_{l,m}^2 - J_{l+1,m}^2 ) \, f_{l,m}^{B} - \right. & \nonumber \\
 \left. \sqrt{(l-2)\,(l-1)} \, J_{l,m} \, J_{l-1,m} \, f_{l-2,m}^{B} - \sqrt{(l+2)\,(l+3)} \, J_{l+1,m} \, J_{l+2,m} \, f_{l+2,m}^{B}\right] & \nonumber
  \end{align}
\end{subequations}
So it seems that we have two different boundary constraints for the~$f^D_{l,m}$. Actually this is not the case, there is no inconsistency. Indeed, eq.~(\ref{eq:BC1}) can be rearranged into
\begin{equation}
\label{eq:BC3}
 \alpha^2 \, \partial_r ( r \, f^D_{l,m}) = \varepsilon_0 \, r \, \tilde{\omega} \, \left[ \sqrt{(l+1)\,(l-1)} \, J_{l,m} \, f^B_{l-1,m} - \sqrt{l\,(l+2)} \, J_{l+1,m} \, f^B_{l+1,m} \right] 
\end{equation}
Inserting this expression into the left hand side of eq.~(\ref{eq:BC2}), we get its right hand side. Therefore, eq.~(\ref{eq:BC2}) is redundant with eq.~(\ref{eq:BC1}), it follows from it. Consequently the correct boundary condition to impose on the $f^D_{l,m}$ is eq.~(\ref{eq:BC3}) and only eq.~(\ref{eq:BC3}). Moreover, because the dipole corresponds to a $m=1$ mode and the problem is linear, we only expect $m=1$ azimuthal modes in the sought solutions.

\subsubsection{Near zone or quasi-static solution}

Before solving numerically the full set of ordinary differential equations, we investigate the near zone solution for $r\ll\rlight$. This is also called the quasi-static regime because it does not contain the electric displacement current. This approximation implies that we can neglect the terms involving $m \, \tilde \omega /c$ in eqs.~(\ref{eq:Helmholtz}). To the lowest order, we find that the magnetic field is given by its static approximation~$f_{1,1}^{B({\rm dip})}$. We therefore look for the first order perturbation in the electric field~$f_{21}^D$, solution of 
\begin{equation}
  \label{eq:StaticfD21a}
  \mathcal{R}_2[f^{D}_{2,1}] = 3 \, \sqrt{\frac{3}{5}} \, \varepsilon_0 \, \frac{\omega}{r} \, f^{B({\rm dip})}_{1,1}
\end{equation}
Written explicitly, we get
\begin{equation}
  \label{eq:StaticfD21b}
  \partial_r(\alpha^2\,\partial_r(r\,f_{2,1}^D)) - \frac{6}{r} \, f_{2,1}^D = -36 \, \frac{\varepsilon_0 \, \mu_0 \, \mu}{4\,\pi} \, \sqrt{\frac{\pi}{5}} \, \frac{a\,c}{R_s^2\,r^2} \, \left[ \ln\alpha^2 + \frac{R_s}{r} + \frac{R_s^2}{2\,r^2} \right]
\end{equation}
which is exactly the same partial differential equation as eq.~(\ref{eq:LaplaceSourcef20}) apart from a constant factor in the inhomogeneous term, in front of $f^{B({\rm dip})}_{1,1}$. Consequently, a particular solution of eq.~(\ref{eq:StaticfD21b}) vanishing at infinity is given by
\begin{equation}
  \label{eq:SolPart21}
  f_{2,1}^{D(p)} = 6\, \sqrt{\frac{\pi}{5}} \, \frac{\varepsilon_0 \, \mu_0 \, \mu}{4\,\pi} \, \frac{a\,c}{R_s^2\,r} \, \left[ \ln\alpha^2 + \frac{R_s}{r} \right]
\end{equation}
The homogeneous solution is again given by eq.~(\ref{eq:DipoleSchwarzf20}). In order to satisfy the boundary condition on the star which is from eq.~(\ref{eq:BC3})
\begin{equation}
  \label{eq:BC}
  \sqrt{5} \, \alpha^2 \, \partial_r(r\,f_{2,1}^{D}) = \sqrt{3} \, \varepsilon_0 \, r \, \tilde{\omega} \, f_{1,1}^{B({\rm dip})}  
\end{equation}
we must set the constant to
\begin{equation}
  \label{eq:Kb2}
  K = - \frac{\varepsilon_0 \, \mu_0 \, \mu}{4\,\pi} \, \frac{C_2}{3\,\alpha_R^2} \, \sqrt{\frac{\pi}{5}} \, \left[ R_s \, R \, \tilde{\omega}_R \, C_1 + \frac{1}{2}\, \, \frac{\omega_R\,R_s^3}{R} \right]
\end{equation}
The full solution reads
\begin{multline}
 f_{2,1}^D = \frac{K}{R_s^2\,r} \, \left[ 6 \, \frac{r^2}{R_s^2} \, \left( 3 - 4 \, \frac{r}{R_s} \right) \, {\rm ln} \left( 1 - \frac{R_s}{r} \right) + 1 + 6 \, \frac{r}{R_s} \, \left( 1 - 4 \, \frac{r}{R_s} \right) \right] \\
 + 6 \, \frac{\varepsilon_0 \, \mu_0 \, \mu}{4\,\pi} \, \sqrt{\frac{\pi}{5}} \, \frac{a\,c}{R_s^2\,r} \, \left[ \ln\alpha^2 + \frac{R_s}{r} \right]
\end{multline}
Taking the value of the constant~$K$ into account, we get
\begin{multline}
   f_{2,1}^D = -3 \, \frac{\varepsilon_0 \, \mu_0 \, \mu}{4\,\pi\,r} \, \sqrt{\frac{\pi}{5}} \left\{ \frac{C_2}{18\,\alpha_R^2} \, \left( \frac{\omega_R\,R_s}{R} + 2 \, C_1 \, \frac{\tilde{\omega_R}\,R}{R_s} \right) \right. \times \\
   \times \left. \left[ 6 \, \frac{r^2}{R_s^2} \, \left( 3 - 4 \, \frac{r}{R_s} \right) \, {\rm ln} \left( 1 - \frac{R_s}{r} \right) + 1 + 6 \, \frac{r}{R_s} \, \left( 1 - 4 \, \frac{r}{R_s} \right) \right] - 2 \, \frac{\omega\,r^3}{R_s^3} \, \left( \ln\alpha^2 + \frac{R_s}{r} \right) \right\}
\end{multline}
This is exactly the same expression as for the aligned rotator, except for a constant factor. Indeed, we have
\begin{equation}
   f_{2,1}^{D({\rm quad})} = - \frac{3}{\sqrt{6}} \, f_{2,0}^{D({\rm quad})}
\end{equation}
The components of the electric field follow then immediately from this remark. In the general case of an oblique rotator with inclination angle~$\chi$ the near zone quasi-static regime of the electric field is given by
\begin{equation}
 \mathbf{D}_1 = \rot ( \cos \chi \, f_{2,0}^{D({\rm quad})} \, \mathbf \Phi_{2,0} + \sin \chi \, f_{2,1}^{D({\rm quad})} \, \mathbf \Phi_{2,1} )
+ \sin \chi \, g_{1,1}^{D({\rm dip})} \, \mathbf \Phi_{1,1}
\end{equation}
Note that we have to add the component related to $g_{1,1}^{D({\rm dip})}$ because it is connected to $f_{1,1}^{B({\rm dip})}$ via eq.~(\ref{eq:gDvsfB}). The components are explicitly
\begin{subequations}
\begin{align}
  \label{eq:Drot3}
  D^{\hat r} & = - \sqrt{\frac{30}{\pi}} \, \frac{f_{2,0}^{D({\rm quad})}}{4\,r} \, ( \cos \chi \, ( 3 \, \cos^2 \vartheta - 1 ) + 3 \, \sin \chi \, \cos\vartheta \, \sin \vartheta \, e^{i\,\varphi} ) \\
  D^{\hat \vartheta} & = \frac{3}{4} \, \sqrt{\frac{5}{6\,\pi}} \, \frac{\alpha}{r} \, \partial_r(r\,f_{2,0}^{D({\rm quad})} ) \, ( 2 \, \cos \chi \, \cos \vartheta \, \sin \vartheta + \sin \chi \, ( \sin^2 \vartheta - \cos^2 \vartheta ) e^{i\,\varphi} ) \\
 & + \frac{1}{2} \, \sqrt{\frac{3}{2\,\pi}} \, \varepsilon_0 \, \frac{\tilde{\omega}}{\alpha} \, f_{1,0}^{B({\rm dip})} \, \sin\chi \, e^{i\,\varphi} \nonumber \\
  D^{\hat \varphi} & = \frac{1}{2} \, \sqrt{\frac{3}{2\,\pi}} \, \left[ - \frac{\sqrt{5}}{2} \, \frac{\alpha}{r} \, \partial_r(r\,f_{2,0}^{D({\rm quad})} ) + \varepsilon_0 \, \frac{\tilde{\omega}}{\alpha} \, f_{1,0}^{B({\rm dip})}\right]  \, \sin \chi \, \cos \vartheta \, i \, e^{i\,\varphi}
\end{align}
\end{subequations}
It is understood that the physical quantities are only the real parts of the above expressions.
These equations are exactly the same as equations (124)-(125)-(126) in \cite{2001MNRAS.322..723R} for the general oblique case, except for a typo in their $E^{\hat \phi}$ component, there should be a minus sign immediately after the first bracket, otherwise $E^{\hat \phi}$ would not vanish on the neutron star surface. It is understood that their $E^{\hat \phi}$ corresponds to our definition of $D^{\hat \varphi}$. In the newtonian limit we find as expected the flat space-time quadrupolar expressions
\begin{subequations}
\begin{align}
  \label{eq:DperpNewt}
  D^{\hat r} & = - \frac{\Omega \, B \, R^5}{r^4} \, ( \cos \chi \, ( 3 \, \cos^2 \vartheta - 1 ) + 3 \, \sin \chi \, \cos\vartheta \, \sin \vartheta \, e^{i\,\varphi} ) \\
  D^{\hat \vartheta} & = - \frac{\Omega \, B \, R^5}{r^4} \, \left[  2 \, \cos \chi \, \cos \vartheta \, \sin \vartheta + \sin \chi \, \left( \sin^2 \vartheta - \cos^2 \vartheta + \frac{r^2}{R^2}\right) \, e^{i\,\varphi} \right]  \\
  D^{\hat \varphi} & = \frac{\Omega \, B \, R^5}{r^4} \, \left( 1 - \frac{r^2}{R^2}\right) \, \sin \chi \, \cos \vartheta \, i \, e^{i\,\varphi}
\end{align}
\end{subequations}
Next we want to look for the solution in the wave zone which leads to a net Poynting flux. We do it by numerical integration of the above mentioned system of partial differential equations, the Helmholtz system with appropriate boundary conditions.

\subsubsection{Numerical solution in whole vacuum}

For the scalar Helmholtz equation, we get into problem applying straightforwardly our expansion into rational Chebyshev functions because the solution oscillates asymptotically. This behaviour cannot be reproduced by the $TL_k$ functions. We therefore supplement these basis functions with an extra function mimicking the right asymptotic behaviour of the solution. We know from the flat space-time expression that it should tend to the spherical Hankel function $h_l^{(1)}(r/\rlight)$. Moreover, we want to impose an asymptotic expansion that tends to only this function. We achieve this by the following expansion of the unknown coefficients~$f_{l,m}^{B/D}$ as
\begin{equation}
 r \, f(r) = \sum_{k=0}^{N_r-2} f_k \, TL_k(y(r)) + f_{N_r-1} \, r \, h_l^{(1)}(r/\rlight)
\end{equation}
and we impose
\begin{equation}
 \lim\limits_{r\to+\infty} \sum_{k=0}^{N_r-2} f_k \, TL_k(y(r)) = 0
\end{equation}
which is simply expressed as
\begin{equation}
 \sum_{k=0}^{N_r-2} f_k = 0
\end{equation}
In this way we get the correct asymptotic behaviour of each coefficient as
\begin{equation}
 \lim\limits_{r\to+\infty} f(r) = f_{N_r-1} \, h_l^{(1)}(r/\rlight)
\end{equation}
We solve numerically the minimal truncated system involving $f_{1,1}^B$ and $f_{2,1}^D$ because of computational resources limitations. From the above discussion, the electromagnetic field is expanded into
\begin{subequations}
 \begin{align}
  \mathbf B & = \rot ( f_{1,1}^B \, \mathbf \Phi_{1,1} ) - i \, \mu_0 \, \frac{\tilde{\omega}}{\alpha} \, f_{2,1}^D \, \mathbf \Phi_{2,1} \\
  \mathbf D & = \rot ( f_{2,1}^2 \, \mathbf \Phi_{2,1} ) + i \, \varepsilon_0 \, \frac{\tilde{\omega}}{\alpha} \, f_{1,1}^B \, \mathbf \Phi_{1,1}
 \end{align}
\end{subequations}
The elliptic problems to be solved are
\begin{subequations}
\label{eq:Perp}
 \begin{align}
 \frac{\alpha^2}{r} \, \frac{\partial}{\partial r} \left( \alpha^2\,\frac{\partial}{\partial r}(r\,f^{B}_{1,1}) \right) - \alpha^2 \, \frac{2}{r^2} \, f^{B}_{1,1} + \frac{(\Omega-\omega)^2}{c^2} \, f^{B}_{1,1} & = 3 \, \sqrt{\frac{3}{5}} \, \mu_0 \, \alpha^2 \, \frac{\omega}{r} \, f^{D}_{2,1} \\
 \frac{\alpha^2}{r} \, \frac{\partial}{\partial r} \left( \alpha^2\,\frac{\partial}{\partial r}(r\,f^{D}_{2,1}) \right) - \alpha^2 \, \frac{6}{r^2} \, f^{D}_{2,1} + \frac{(\Omega-\omega)^2}{c^2} \, f^{D}_{2,1} & = 3 \, \sqrt{\frac{3}{5}} \, \varepsilon_0 \, \alpha^2 \, \frac{\omega}{r} \, f^{B}_{1,1}
 \end{align}
\end{subequations}
and the boundary condition is the same as in the quasi-static regime, eq.~(\ref{eq:BC}). In the asymptotic limit of very large distances, we know that the solution relaxes to the Deutsch field, therefore
\begin{subequations}
 \begin{align}
  \lim\limits_{r\to+\infty} f_{1,1}^B & = f_{1,1}^{B(\infty)} \, h_1^{(1)} \left(\frac{r}{\rlight}\right) \\
  \lim\limits_{r\to+\infty} f_{2,1}^D & = f_{2,1}^{D(\infty)} \, h_2^{(1)} \left(\frac{r}{\rlight}\right)
 \end{align}
\end{subequations}
where $f_{1,1}^{B(\infty)}$ and $f_{2,1}^{D(\infty)}$ are two constants derived from the numerical solution of eqs.(\ref{eq:Perp}), actually corresponding to the last term $f_{N_r-1}$ in the expansion.

\begin{figure}
  \centering
 \includegraphics[width=0.5\textwidth]{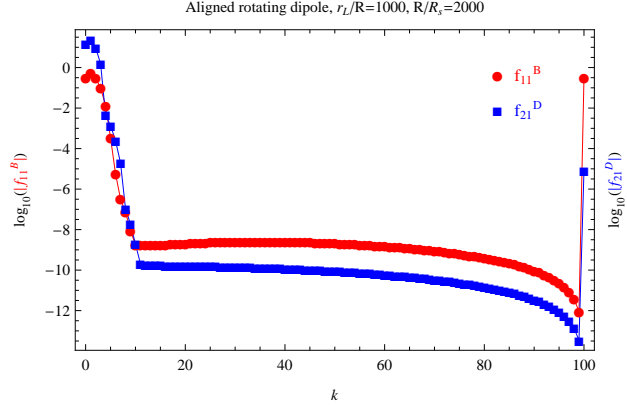}
  \caption{Absolute value of the coefficients of the rational Chebyshev expansion of the magnetic field and electric field functions $f_{1,1}^{B}$ and $f_{2,1}^{D}$ for the perpendicular rotating dipole for $R=2000\,\Rs$ and $\rlight=1000\,R$. The large value of the last coefficient in the expansion corresponds to the asymptotic behavior related to the spherical Hankel functions.}
\label{fig:u_perp_rot_1_r1000_rs2000}
\end{figure}

\begin{figure}
  \centering
 \includegraphics[width=0.5\textwidth]{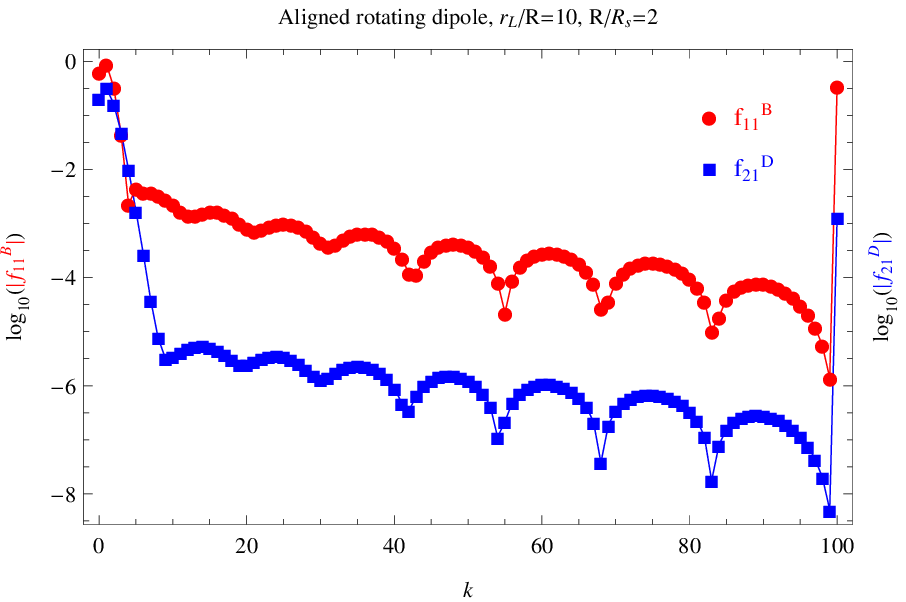}
  \caption{Same as fig.~\ref{fig:u_perp_rot_1_r1000_rs2000} but for $R=2\,\Rs$ and $\rlight=10\,R$.}
\label{fig:u_perp_rot_1_r10_rs2}
\end{figure}

Two examples of the coefficients obtained by this procedure for $f_{1,1}^{B}$ and $f_{2,1}^{D}$ are shown in the non relativistic limit with $R=2000\,\Rs$ and $\rlight=1000\,R$, fig.\ref{fig:u_perp_rot_1_r1000_rs2000}, and in the extreme relativistic limit with $R=2\,\Rs$ and $\rlight=10\,R$, fig.\ref{fig:u_perp_rot_1_r10_rs2}. The convergence of the first few coefficients is fast but after number ten or so, the decrease in the magnitude of the coefficient becomes rather weak. This is probably due to the asymptotic expression we choose as spherical Hankel functions, those useful in flat space-time. Switching to more accurate asymptotic behaviour in a Schwarzschild background metric would certainly help to improve the convergence but such functions do not (yet) exist in the literature. The presence of the lapse function~$\alpha$ make the convergence to spherical Hankel functions only first order.
To conclude, we compute the Poynting flux at infinity and compare it with the flat space-time value obtained from the point magnetic dipole losses. The Poynting vector is given by $\mathbf{S} = \mathbf{E} \wedge \mathbf{H}$ but asymptotically the electric field $\varepsilon_0 \, \mathbf E$ tends towards $\mathbf D$ and the magnetic field $\mu_0 \, \mathbf H$ tends towards $\mathbf B$. In order to get the spin-down of the neutron star, we only need the radial component of the Poynting vector such that $S_{\rm r} = c^2 \, ( D_\vartheta \, B_\varphi - D_\varphi \, B_\vartheta)$. We already know that the Poynting flux for the perpendicular rotator in flat space-time is given by
\begin{equation}
 L_{\rm sd}^{\rm flat} = \frac{\mu_0 \, c}{6\,\pi} \, \frac{\mu^2}{\rlight^4} = \frac{8\,\pi}{3\,\mu_0\,c^3} \, \Omega^4\,B^2\,R^6
\end{equation}
For the Deutsch field, this flux is
\begin{equation}
 L_{\rm sd}^{\rm Deustch} = \frac{4}{5} \, \frac{45 - 3\,x^4 + 2\,x^6}{(1 + x^2)\,(36 - 3\,x^4 + x^6)} \, L_{\rm sd}^{\rm flat} \approx ( 1 - x^2 ) \, L_{\rm sd}^{\rm flat}
\end{equation}
where $x=R/\rlight$ and the approximation is valid for $x\ll1$. In the general-relativistic case we have
\begin{equation}
 L_{\rm sd}^{\rm gr} = \int S_{\rm r} \, r^2 \, d\Omega = \frac{1}{2} \, (|f_{1,1}^{B(\infty)}|^2 + |f_{2,1}^{D(\infty)}|^2 )
\end{equation}
For comparison between general-relativistic situation and flat space-time we compute the normalized flux as
\begin{equation}
\label{eq:Poynting}
 \frac{L_{\rm sd}^{\rm gr}}{L_{\rm sd}^{\rm flat}} = 3 \, \pi \, (|f_{1,1}^{B(\infty)}|^2 + |f_{2,1}^{D(\infty)}|^2 )
\end{equation}
This expression does not take into account neither the magnetic field amplification as measured at the surface of the neutron star no the gravitational redshift of the rotation frequency. These can be deduced analytically from the lapse function~$\alpha$ and from the expression of the magnetic field in curved space-time, see \cite{2004MNRAS.352.1161R}. The ratio in eq.~(\ref{eq:Poynting}) is shown in table~\ref{tab:FluxPoynting}. With our choice of normalization according to the point dipole formula, we don't notice any significant change in the Poynting flux, modulo gravitational redshift and field amplification. The difference is at most 15\%, which is much less than the previously mentioned effects. Therefore, the order of magnitude given by \cite{2004MNRAS.352.1161R} is actually a good estimate of the magnetic dipole losses of an orthogonal rotator in general relativity.
\begin{table}
\begin{center}
\begin{tabular}{rrrrr}
\hline
$R/\Rs$ & $\rlight/R$ & Point & Deutsch & General \\ 
 &  & dipole & field & relativity \\ 
\hline
\hline
2000 & 1000 & 1 & 0.9999 & 1.0088 \\ 
2000 &   10 & 1 & 0.9901 & 0.9645 \\ 
   2 & 1000 & 1 & 0.9999 & 1.0226 \\
   2 &   10 & 1 & 0.9901 & 1.1570 \\
\hline
\end{tabular}
\end{center}
 \caption{Normalized Poynting flux for the general-relativistic perpendicular rotator compared to the expectation from the point dipole losses. Neither gravitational redshift nor magnetic field amplification are taken into account here. This should emphasize the effect of frame-dragging only.}
 \label{tab:FluxPoynting}
\end{table}

\section{CONCLUSION}
\label{sec:Conclusion}

In this paper, we showed how to look for a systematic solution to the stationary Maxwell equations in the background space-time of a slowly rotating neutron star following the 3+1 foliation and an expansion of the unknown electromagnetic field onto vector spherical harmonics. We obtained numerical solutions of high accuracy for the aligned rotator and less accurate for the orthogonal rotator. We hope that these results will serve as a benchmark for general-relativistic codes solving the electromagnetic field in a static background metric like for instance pulsar and black hole magnetospheres. The orthogonal rotator could still benefit from some improvements by replacing the asymptotic spherical Hankel functions by more precise functions which take into account at least to first order the perturbation in the metric induced by the presence of the mass~$M$. This would lead to more rapid convergence of the solution but those analytical solutions do not exist.

A next step would be to solve the time dependent Maxwell equations instead of looking for solutions to the boundary value problem, especially difficult to handle with high accuracy for an orthogonal rotator. This could improve the estimate of the magnetic dipole losses in a curved space-time.

A further step to this work will be to include a force-free plasma surrounding the neutron star in order to compute the pulsar force-free magnetosphere in the general-relativistic case. The same technique could be useful for the black hole magnetosphere. However, because of the non linearity implied by the force-free current, it is impossible to solve the system semi-analytically as we did here. Computing the force-free magnetosphere requires numerical simulations. This is the subject of a forthcoming paper in which we will describe a time dependent pseudo-spectral code using the vector spherical harmonics expansion to solve Maxwell equations in a curved vacuum space-time.

\section*{Acknowledgments}
I would like to thank \'Eric Gourgoulhon and Serguei Komissarov for helpful discussions.


\appendix 

\onecolumn

\section{3+1 metric}
\label{app:metric}

We give the explicit expressions for the metric and the electromagnetic field tensor using the Landau-Lifschitz signature $(+,-,-,-)$ \cite{LandauLifchitzTome2}. The metric decomposed into time and space components reads then
\begin{equation}
 g_{ik} = 
 \begin{pmatrix}
  \alpha^2-\beta^2 & - \beta_b \\
  -\beta_a & -\gamma_{ab}
 \end{pmatrix} =
 \begin{pmatrix}
  \alpha^2-\beta^2 & - \beta_1 & - \beta_2 & - \beta_3  \\
  -\beta_1 & -\gamma_{11} & -\gamma_{12} & -\gamma_{13} \\
  -\beta_2 & -\gamma_{21} & -\gamma_{22} & -\gamma_{23} \\
  -\beta_3 & -\gamma_{31} & -\gamma_{32} & -\gamma_{33}
 \end{pmatrix}
\end{equation}
where $\alpha$ is the lapse function, $\beta^a$ the shift vector and $\beta^2 = \beta^a \, \beta_a$. The spatial metric is simply given by~$\gamma_{ab} = - g_{ab}$ and the
inverse metric by
\begin{eqnarray}
 g^{ik} & = &
 \begin{pmatrix}
  1/\alpha^2 & - \beta^b / \alpha^2 \\
  - \beta^a / \alpha^2 & -\gamma^{ab} + \beta^a \, \beta^b / \alpha^2
 \end{pmatrix} \\
 & = &
 \begin{pmatrix}
  1/\alpha^2 & -\beta^1/\alpha^2 & -\beta^2/\alpha^2 & -\beta^3/\alpha^2 \\
  -\beta^1/\alpha^2 & -\gamma^{11} + \beta^1 \, \beta^1 / \alpha^2 & -\gamma^{12} + \beta^1 \, \beta^2/\alpha^2 & -\gamma^{13} + \beta^1 \, \beta^3/\alpha^2 \\
  -\beta^2/\alpha^2 & -\gamma^{21} + \beta^2 \, \beta^1 / \alpha^2 & -\gamma^{22} + \beta^2 \, \beta^2 /\alpha^2 & -\gamma^{23} + \beta^2 \, \beta^3/\alpha^2 \\
  -\beta^3/\alpha^2 & -\gamma^{31} + \beta^3 \, \beta^1 / \alpha^2 & -\gamma^{32} + \beta^3 \, \beta^2/\alpha^2 & -\gamma^{33} + \beta^3 \, \beta^3/\alpha^2
 \end{pmatrix}
\end{eqnarray}
The contravariant components of the electromagnetic field tensor expressed with the fields $(\mathbf{D}, \mathbf{H})$ are
\begin{equation}
 \label{eq:TenseurChampEMContra1}
 F^{ik} = \frac{1}{\sqrt{-g}} 
\begin{pmatrix} 
  0 & - \sqrt{\gamma} \, D^1/\varepsilon_0\,c & - \sqrt{\gamma} \, D^2/\varepsilon_0\,c & - \sqrt{\gamma} \, D^3/\varepsilon_0\,c \\
 \sqrt{\gamma} \, D^1/\varepsilon_0\,c &      0 & - \mu_0 \, H_3 &   \mu_0 \, H_2 \\
 \sqrt{\gamma} \, D^2/\varepsilon_0\,c &   \mu_0 \, H_3 &      0 & - \mu_0 \, H_1 \\
 \sqrt{\gamma} \, D^3/\varepsilon_0\,c & - \mu_0 \, H_2 &    \mu_0 \, H_1 &    0
\end{pmatrix}
\end{equation}
and its dual expressed with the fields $(\mathbf{E}, \mathbf{B})$ are
\begin{equation}
 \label{eq:TenseurChampEMContra2}
 {^*F}^{ik} = \frac{1}{\sqrt{-g}} 
\begin{pmatrix} 
  0 & - \sqrt{\gamma} \, B^1 & - \sqrt{\gamma} \, B^2 & - \sqrt{\gamma} \, B^3 \\
 \sqrt{\gamma} \, B^1 &       0 &   E_3/c & - E_2/c \\
 \sqrt{\gamma} \, B^2 & - E_3/c &       0 &   E_1/c \\
 \sqrt{\gamma} \, B^3 &   E_2/c & - E_1/c &    0
\end{pmatrix}
\end{equation}
The covariant components of the electromagnetic field tensor expressed with the fields $(\mathbf{E}, \mathbf{B})$ are
\begin{equation}
 \label{eq:TenseurChampEMContra3}
 F_{ik} =
\begin{pmatrix} 
  0 & E_1/c & E_2/c & E_3/c \\
 - E_1/c &      0 & - \sqrt{\gamma} \, B^3 &  \sqrt{\gamma}\,B^2 \\
 - E_2/c &   \sqrt{\gamma} \, B^3 &      0 & - \sqrt{\gamma} \, B^1 \\
 - E_3/c & - \sqrt{\gamma} \, B^2 &  \sqrt{\gamma} \, B^1 &    0
\end{pmatrix}
\end{equation}
and for its dual with respect to $(\mathbf{D}, \mathbf{H})$ are
\begin{equation}
 \label{eq:TenseurChampEMContra4}
 {^*F}_{ik} =
\begin{pmatrix} 
  0 & \mu_0 \, H_1 & \mu_0 \, H_2 & \mu_0 \, H_3 \\
 - \mu_0 \, H_1 &      0 & \sqrt{\gamma} \, D^3/\varepsilon_0\,c & - \sqrt{\gamma}\,D^2 /\varepsilon_0\,c \\
 - \mu_0 \, H_2 & - \sqrt{\gamma} \, D^3/\varepsilon_0\,c &      0 & \sqrt{\gamma} \, D^1/\varepsilon_0\,c \\
 - \mu_0 \, H_3 &   \sqrt{\gamma} \, D^2/\varepsilon_0\,c & - \sqrt{\gamma} \, D^1/\varepsilon_0\,c &    0
\end{pmatrix}
\end{equation}

\section{Differential operators in curved space}
\label{app:operateur}

The metric of a slowly rotating neutron star remains very close to the usual flat space, except for the radial direction. Indeed the spatial metric is diagonal such that
\begin{equation}
  \label{eq:Metric3D}
  \gamma_{ab} =
  \begin{pmatrix}
    \alpha^{-2} & 0 & 0 \\
    0 & r^2 & 0 \\
    0 & 0 & r^2 \sin^2\vartheta
  \end{pmatrix}
\end{equation}
For any scalar field~$f$, the gradient and Laplacian are respectively
\begin{subequations}
 \begin{align}
  \grad f & = e_{\hat r} \, \alpha \, \partial_r f + e_{\hat \vartheta} \, \frac{1}{r} \, \partial_\vartheta f + e_{\hat \varphi} \, \frac{1}{r\,\sin\vartheta} \, \partial_\varphi f \\
  \Delta f & = \frac{\alpha}{r^2} \, \partial_r ( \alpha \, r^2 \, \partial_r f ) + \frac{1}{r^2\,\sin\vartheta} \, \partial_\vartheta ( \sin\vartheta \, \partial_\vartheta f ) + \frac{1}{r^2\,\sin^2\vartheta} \, \partial^2_\varphi f
 \end{align}
\end{subequations}
The physical components of a vector are depicted by hatted indexes. The differential vector operators are then for the divergence and the curl
\begin{subequations}
 \begin{align}
  \label{eq:Div}
  \divg \mathbf B & = \frac{\alpha}{r^2} \, \partial_r(r^2\,B^{\hat r}) + \frac{1}{r\,\sin\vartheta} \, \partial_\vartheta(\sin\vartheta\,B^{\hat \vartheta}) + \frac{1}{r\,\sin\vartheta} \, \partial_\varphi\,B^{\hat \varphi} \\
  (\rot \mathbf B)^{\hat r} & = \frac{1}{r\,\sin\vartheta} \, \left[ \partial_\vartheta(\sin\vartheta\,B^{\hat \varphi}) - \partial_\varphi\,B^{\hat \vartheta} \right] \\
  (\rot \mathbf B)^{\hat \vartheta} & = \frac{1}{r\,\sin\vartheta} \, \partial_\varphi\,B^{\hat r} - \frac{\alpha}{r} \, \partial_r(r\,B^{\hat \varphi}) \\
  (\rot \mathbf B)^{\hat \varphi} & = \frac{\alpha}{r} \, \partial_r(r\,B^{\hat \vartheta}) - \frac{1}{r} \, \partial_\vartheta\,B^{\hat r}
 \end{align}
\end{subequations}
These expressions are very similar to their flat space equivalent, except for the replacement of the radial derivative $\partial_r$ by $\alpha\,\partial_r$ in each term. The transverse part of the spatial metric $\gamma_{ab}$ with $(a,b)\in(\vartheta,\varphi)$ is exactly the same as for the flat space. Because the flat space vector spherical harmonics (VSH) lie only in this transverse sub-space, it is straightforward to extend these VSH to the special metric Eq.~(\ref{eq:Metric3D}) as shown in the next paragraph.

\section{Vector spherical harmonics in curved space}
\label{app:HSV}

We generalize the vector spherical harmonics (VSH) introduced in \cite{2012MNRAS.424..605P} to a three-dimensional curved space. The three sets of vector spherical harmonics we use are defined by
\begin{subequations}
 \begin{align}
  \label{eq:Ylm_vect_def}
  \mathbf{Y}_{l,m}    & = Y_{l,m} \, \mathbf{e}_{\rm r} \\
  \label{eq:Psilm_vect_def}
  \mathbf{\Psi}_{l,m} & = \frac{r}{\sqrt{l\,(l+1)}} \, \grad Y_{l,m} \\
  \label{eq:Philm_vect_def}
  \mathbf{\Phi}_{l,m} & = \frac{\mathbf{r}}{\sqrt{l\,(l+1)}} \, \times \grad Y_{l,m}
 \end{align}
\end{subequations}
Any smooth three-dimensional vector field~$\mathbf{E}$ admits an
expansion onto these vectors according to
\begin{equation}
  \label{eq:Decomposition_HSV_general}
  \mathbf{E}(r,\vartheta,\varphi) = \sum_{l=0}^\infty\sum_{m=-l}^l
  \left(E^r_{l,m}(r)\mathbf{Y}_{l,m}+E^{(1)}_{l,m}(r)\mathbf{\Psi}_{l,m}+
    E^{(2)}_{l,m}(r)\mathbf{\Phi}_{l,m}\right)
\end{equation}

\subsection{Properties}

The vector spherical harmonics share some useful properties with respect to their spatial derivatives. First, assume a 3D scalar field~$\phi$ expanded onto the scalar spherical harmonics such that
\begin{eqnarray}
  \label{eq:MultipoleScalaire}
  \phi(r,\vartheta,\varphi) & = & \sum_{l=0}^\infty \sum_{m=-l}^l \phi_{l,m}(r) \, Y_{l,m}(\vartheta,\varphi)
\end{eqnarray}
Then, its gradient expanded onto the VSH becomes
\begin{equation}
  \label{eq:Gradient}
  \grad \phi = \sum_{l=0}^\infty \sum_{m=-l}^l 
  \left(\alpha \, \frac{\partial\phi_{l,m}}{\partial r} \mathbf{Y}_{l,m} +
    \frac{\sqrt{l\,(l+1)}}{r} \, \phi_{l,m} \, \mathbf{\Psi}_{l,m}\right) 
\end{equation}
The action of the same gradient on the VSH gives the divergence of any vector field~$\mathbf{E}$ as
\begin{equation}
\divg\mathbf{E} = \sum_{l=0}^\infty \sum_{m=-l}^l 
\left( \frac{\alpha}{r^2} \, \frac{\partial}{\partial r} ( r^2 \, E^r_{l,m} ) - 
  \frac{\sqrt{l(l+1)}}{r}E^{(1)}_{l,m}\right)Y_{l,m}
\end{equation}
and for the curl
\begin{multline}
\rot \mathbf{E} = \sum_{l=0}^\infty \sum_{m=-l}^l \left[ - \frac{\sqrt{l(l+1)}}{r} \, E^{(2)}_{l,m} \, \mathbf{Y}_{l,m} - \frac{\alpha}{r} \, \frac{\partial}{\partial r} (r\,E^{(2)}_{l,m}) \mathbf{\Psi}_{l,m} \right. \\
  + \left. \left(- \frac{\sqrt{l(l+1)}}{r} \, E^r_{l,m} + \frac{\alpha}{r} \, \frac{\partial}{\partial r}(r\,E^{(1)}_{l,m}) \right) \, \mathbf{\Phi}_{l,m} \right]
\end{multline}
For each component taken individually, we find for the divergence
\begin{subequations}
 \begin{align}
  \label{eq:DivY}
  \divg \left(f(r) \, \mathbf{Y}_{l,m} \right) & = \frac{\alpha}{r^2} \frac{\partial}{\partial r} \left( r^2 \, f \right) \, Y_{l,m} \\
  \divg \left(f(r) \, \mathbf{\Psi}_{l,m}\right) & = - \frac{\sqrt{l(l+1)}}{r} \, f \, Y_{l,m} \\
  \divg\left(f(r)\mathbf{\Phi}_{l,m}\right) & = 0
 \end{align}
\end{subequations}
and for the curl
\begin{subequations}
 \begin{align}
  \label{eq:Rot}
  \rot \left(f(r) \, \mathbf{Y}_{l,m}\right) & =- \frac{\sqrt{l\,(l+1)}}{r} \, f \, \mathbf{\Phi}_{l,m} \\
  \rot \left(f(r)\mathbf{\Psi}_{l,m}\right) & = \frac{\alpha}{r} \, \frac{\partial}{\partial r}(r\,f) \mathbf{\Phi}_{l,m} \\
  \rot \left(f(r)\mathbf{\Phi}_{l,m}\right) & = - \frac{\sqrt{l(l+1)}}{r} \, f \, \mathbf{Y}_{l,m} - \frac{\alpha}{r} \, \frac{\partial}{\partial r}(r\,f) \, \mathbf{\Psi}_{l,m}
 \end{align}
\end{subequations}
Finally, define the radial differential operator~$\mathcal{D}_l$ by
\begin{equation}
  \label{eq:MatchalD}
  \mathcal{D}_l[f] = \frac{\alpha}{r} \, \frac{d}{dr} \left( \alpha \, \frac{d}{dr} (r\,f) \right) - \frac{l(l+1)}{r^2} \, f = \frac{\alpha}{r^2} \, \frac{d}{dr} \left( \alpha \, r^2 \, \frac{df}{dr}  \right) - \frac{l(l+1)}{r^2} \, f
\end{equation}
The VSH noted $\mathbf{\Phi}_{l,m}$ are eigenvectors in the linear
algebra meaning, of the vector Laplacian operator~$\Delta$. Indeed, it
is straightforward to show that
\begin{equation}
  \label{eq:nabl}
  \Delta [ f(r) \, \mathbf{\Phi}_{l,m} ] = \mathcal{D}_l[f] \, \mathbf{\Phi}_{l,m}
\end{equation}
It is thus an extension of the properties of the scalar spherical harmonics to the realm of 3D vectors. Another useful relation is
\begin{equation}
\label{eq:RotRotPhi}
  \rot ( \alpha \, \rot \left(f(r) \, \mathbf{\Phi}_{l,m} \right) ) = - \alpha \, \mathcal{R}_l[f] \, \mathbf{\Phi}_{l,m}
\end{equation}
where we introduced the operator
\begin{equation}
 \label{eq:rdiff}
  \mathcal{R}_l[f] \equiv \left[ \frac{1}{r} \, \frac{\partial}{\partial r} \left( \alpha^2\,\frac{\partial}{\partial r}(r\,f) \right) - \frac{l(l+1)}{r^2} \, f \right]
\end{equation}

\subsection{Expansion of a vector field onto VSH}

From the above discussion, the components of any vector field can be computed according to the three underlying equalities
\begin{subequations}
 \begin{align}
 E_r & = \sum_{l,m} E_{l,m}^r \, Y_{l,m} \\
 \mathbf{\nabla}_{\vartheta,\varphi} \cdot \mathbf E & = \sum_{l,m} - \frac{\sqrt{l(l+1)}}{r} \, E_{l,m}^{(1)} \, Y_{l,m} \\
 \rot \mathbf E \cdot \er & = \sum_{l,m} - \frac{\sqrt{l(l+1)}}{r} \, E_{l,m}^{(2)} \, Y_{l,m}
 \end{align}
\end{subequations}
where $\mathbf{\nabla}_{\vartheta,\varphi} \cdot \mathbf E$ means taking only the angular
part of the divergence.  More explicitly, from the definition of the
differential operators, we get
\begin{subequations}
 \begin{align}
 E_r & = \sum_{l,m} E_{l,m}^r \, Y_{l,m} \\
 \frac{1}{\sin\vartheta} \, \partial_\vartheta (\sin\vartheta \, E_\vartheta ) +  \frac{1}{\sin\vartheta} \, \partial_\varphi \, E_\varphi & = \sum_{l,m} - \sqrt{l(l+1)} \, E_{l,m}^{(1)} \, Y_{l,m} \\
 \frac{1}{\sin\vartheta} \, \partial_\vartheta (\sin\vartheta \, E_\varphi ) -  \frac{1}{\sin\vartheta} \, \partial_\varphi \, E_\vartheta & = \sum_{l,m} - \sqrt{l(l+1)} \, E_{l,m}^{(2)} \, Y_{l,m}
 \end{align}
\end{subequations}
Finding the components of~$\mathbf{E}$ is therefore equivalent to finding
the expansion coefficients of three scalar fields onto the scalar
spherical harmonics. This procedure works for any vector
field. However, the magnetic field being divergencelessness, only two of
the three components are independent. It is therefore judicious to
deal properly with those kind of fields by analytically enforcing the
condition on the divergence as explained below.

\subsection{Expansion of a divergencelessness vector field}

Any divergencelessness vector field is efficiently developed onto the
vector spherical harmonic \textit{orthonormal basis}. This will be the
case for the magnetic field and the electric field in our algorithm.

Assume that the vector field~$\mathbf{V}$ is divergencelessness. It is helpful to
introduce two scalar functions $f_{l,m}(r,t)$ and $g_{l,m}(r,t)$ such
that the decomposition immediately implies the property of
divergencelessness field. This is achieved by writing
\begin{equation}
  \label{eq:Decomposition_HSV_div_0}
  \mathbf{V}(r,\vartheta,\varphi,t) = \sum_{l=1}^\infty\sum_{m=-l}^l
  \left( \rot [f_{l,m}(r,t) \, \mathbf{\Phi}_{l,m}] + 
    g_{l,m}(r,t) \, \mathbf{\Phi}_{l,m} \right)
\end{equation}
This expression automatically and \textit{analytically} enforces the
condition $\divg \mathbf{V} = 0$.  Let us quickly draw the way to
compute these functions.  The transformation from the spherical
components to the functions $(f_{l,m},g_{l,m})$ is given by
\begin{subequations}
 \begin{align}
 \mathbf{V} \cdot \er & = \sum_{l=1}^\infty\sum_{m=-l}^l - \frac{\sqrt{l\,(l+1)}}{r} \, f_{l,m} \, Y_{l,m} \\
 (\rot \mathbf{V}) \cdot \er & = \sum_{l=1}^\infty\sum_{m=-l}^l - \frac{\sqrt{l\,(l+1)}}{r} \, g_{l,m} \, Y_{l,m}
 \end{align}
\end{subequations}
Thus, it is sufficient to expand again the radial component of the
vector and its curl onto scalar spherical harmonics. We get
\begin{subequations}
 \begin{align}
 r \, V_r & = \sum_{l,m} - \sqrt{l\,(l+1)} \, f_{l,m} \, Y_{l,m} \\
 \frac{1}{\sin\vartheta} \, \partial_\vartheta (\sin\vartheta \, V_\varphi ) -  \frac{1}{\sin\vartheta} \, \partial_\varphi \, V_\vartheta & = \sum_{l,m} - \sqrt{l(l+1)} \, g_{l,m} \, Y_{l,m}
 \end{align}
\end{subequations}
The functions $(f_{l,m},g_{l,m})$ are related to the general expansion
Eq.~(\ref{eq:Decomposition_HSV_general}) by
\begin{subequations}
 \begin{align}
 V_{l,m}^r & = - \frac{\sqrt{l\,(l+1)}}{r} \, f_{l,m} \\
 V_{l,m}^{(1)} & = - \frac{\alpha}{r} \, \partial_r(r\,f_{l,m}) \\
 V_{l,m}^{(2)} & = g_{l,m}
 \end{align}
\end{subequations}

\subsection{Useful identities for frame dragging effects}
\label{sec:framedragging}

In our 3+1 formulation of Maxwell equations in curved space, the frame dragging effects are included in the constitutive relations Eq.~(\ref{eq:ConstitutiveE}), (\ref{eq:ConstitutiveH}), i.e. the cross product of two divergencelessness vector fields $\mathbf{\beta}$ and $\mathbf{D}$ or $\mathbf{\beta}$ and $\mathbf{B}$.

From the definition of the VSH and the shift vector we get
\begin{equation}
 \mathbf{\beta} \times (f\,\mathbf{\Phi}_{l,m}) = -i\,\frac{m}{\sqrt{l(l+1)}} \, \frac{\omega\,r}{c} \, f \, \mathbf{Y}_{l,m}
\end{equation}
and therefore for the curl
\begin{equation}
\label{eq:RotBETAVHS}
 \rot (\mathbf{\beta} \times (f\,\mathbf{\Phi}_{l,m}) ) = i \, m \, \frac{\omega}{c} \, f \, \mathbf{\Phi}_{l,m}
\end{equation}
The second useful set of identities involves
\begin{equation}
\label{eq:BETAVHS}
 \mathbf{\beta} \times \rot (f\,\mathbf{\Phi}_{l,m}) = \frac{\omega}{c} \, \sin \vartheta \, \left[ \sqrt{l(l+1)} \, f \, Y_{l,m} \, \etheta - \frac{\alpha}{\sqrt{l(l+1)}} \partial_r(r\,f) \, \partial_\vartheta Y_{l,m} \, \er \right]
\end{equation}
Applying straightforward algebra using the VSH definitions and the scalar harmonics eigenfunction properties, we get
\begin{multline}
\label{eq:RotBETARotVHS}
 \rot (\mathbf{\beta} \times \rot (f\,\mathbf{\Phi}_{l,m}) ) = i \, m \, \frac{\omega\,r^3}{c} \, \rot \left( \frac{f}{r^3}\,\mathbf{\Phi}_{l,m} \right) \\
+ 3 \, \alpha \, \frac{\omega}{c\,r} \, f \, \left[ \sqrt{l\,(l+2)} \, J_{l+1,m} \, \mathbf{\Phi}_{l+1,m} - \sqrt{(l+1)(l-1)} \, J_{l,m} \, \mathbf{\Phi}_{l-1,m} \right]
\end{multline}
We give explicit expressions for the first few modes $l=1,2,3$ for the aligned rotator $m=0$
\begin{subequations}
\begin{align}
 \rot (\mathbf{\beta} \times (f\,\mathbf{\Phi}_{1,0}) ) & = 0 \\
 \rot (\mathbf{\beta} \times (f\,\mathbf{\Phi}_{2,0}) ) & = 0 \\
 \rot (\mathbf{\beta} \times (f\,\mathbf{\Phi}_{3,0}) ) & = 0 \\
 \rot (\mathbf{\beta} \times \rot (f\,\mathbf{\Phi}_{1,0}) ) & = \frac{\alpha\,\omega}{c\,r} \, f \, \frac{6}{\sqrt{5}} \, \mathbf{\Phi}_{20} \\
 \rot (\mathbf{\beta} \times \rot (f\,\mathbf{\Phi}_{20}) ) & = \frac{\alpha\,\omega}{c\,r} \, f \, \left( - \frac{6}{\sqrt{5}} \, \mathbf{\Phi}_{1,0} + 18 \, \sqrt{\frac{2}{35}} \, \mathbf{\Phi}_{30} \right) \\
 \rot (\mathbf{\beta} \times \rot (f\,\mathbf{\Phi}_{30}) ) & = \frac{\alpha\,\omega}{c\,r} \, f \, \left( - 18 \, \sqrt{\frac{2}{35}} \, \mathbf{\Phi}_{20} + 4 \, \sqrt{\frac{15}{7}} \, \mathbf{\Phi}_{40} \right)
\end{align} 
\end{subequations}
and for the perpendicular rotator $m=1$
\begin{subequations}
\begin{align}
 \rot (\mathbf{\beta} \times (f\,\mathbf{\Phi}_{11}) ) & = i \, \frac{\omega}{c} \, f \, \mathbf{\Phi}_{11} \\
 \rot (\mathbf{\beta} \times (f\,\mathbf{\Phi}_{21}) ) & = i \, \frac{\omega}{c} \, f \, \mathbf{\Phi}_{21} \\
 \rot (\mathbf{\beta} \times (f\,\mathbf{\Phi}_{31}) ) & = i \, \frac{\omega}{c} \, f \, \mathbf{\Phi}_{31} \\
 \rot (\mathbf{\beta} \times \rot (f\,\mathbf{\Phi}_{11}) ) & = \frac{\omega}{c\,r} \, \left( -\,i\,\sqrt{2} \, f \, \mathbf{Y}_{11} - i\, r^3 \, \alpha \, \partial_r \left( \frac{f}{r^2}\right) \, \mathbf{\Psi}_{11} + 3 \, \sqrt{\frac{3}{5}} \, \alpha \, f \, \mathbf{\Phi}_{21} \right) \\
 & = \frac{\omega\,r^3}{c} \, \left( i \, \rot \left( \frac{f}{r^3} \, \mathbf{\Phi}_{11} \right) + 3 \, \sqrt{\frac{3}{5}} \, \frac{\alpha}{r^4} \, f \, \mathbf{\Phi}_{21} \right) \\
 \rot (\mathbf{\beta} \times \rot (f\,\mathbf{\Phi}_{21}) ) & = \frac{\omega}{c\,r} \, \left( -\,i\,\sqrt{6} \, f \, \mathbf{Y}_{21}  - i\,r^3 \, \alpha\,  \partial_r \left( \frac{f}{r^2}\right) \,  \mathbf{\Psi}_{21} - 3 \, \sqrt{\frac{3}{5}} \, \alpha \, f \, \mathbf{\Phi}_{11} + \frac{24}{\sqrt{35}} \, \alpha \, f \, \mathbf{\Phi}_{31} \right) \\
 & = \frac{\omega\,r^3}{c} \, \left( i \, \rot \left( \frac{f}{r^3} \, \mathbf{\Phi}_{21} \right) - 3 \, \sqrt{\frac{3}{5}} \, \frac{\alpha}{r^4} \, f \, \mathbf{\Phi}_{11} + \frac{24}{\sqrt{35}} \, \frac{\alpha}{r^4} \, f \, \mathbf{\Phi}_{31} \right) \\
 \rot (\mathbf{\beta} \times \rot (f\,\mathbf{\Phi}_{31}) ) & = \frac{\omega}{c\,r} \, \left( -2\,i\,\sqrt{3} \, f \, \mathbf{Y}_{31} - i\, r^3 \, \alpha\, \partial_r \left( \frac{f}{r^2}\right) \, \mathbf{\Psi}_{31} - \frac{24}{\sqrt{35}} \, \alpha \, f \, \mathbf{\Phi}_{21} + \frac{15}{\sqrt{7}} \, \alpha \, f \, \mathbf{\Phi}_{41} \right) \\
 & = \frac{\omega\,r^3}{c} \, \left( i \, \rot \left( \frac{f}{r^3} \, \mathbf{\Phi}_{31} \right) - \frac{24}{\sqrt{35}} \, \frac{\alpha}{r^4} \, f \, \mathbf{\Phi}_{21} + \frac{15}{\sqrt{7}} \, \frac{\alpha}{r^4} \, f \, \mathbf{\Phi}_{41} \right)
\end{align}
\end{subequations}

\label{lastpage}

\end{document}